\documentclass[a4,usenatbib]{mn2e}

\usepackage{epsfig}
\usepackage{amsmath}
\usepackage{natbib}

\def\apj{ApJ}
\def\aap{A\&A}
\def\apjl{ApJL}
\def\apjs{ApJS}
\def\mnras{MNRAS}
\def\aj{AJ}
\def\araa{ARA\&A}
\def\nat{Nature}

\def\linexp{{lin-exp}}
\def\ttr{{t_{\rm trans}}}
\def\gyr{\,{\rm Gyr}}

\newcommand{\hmpc}{h^{-1}{\rm Mpc}}

\newcommand{\Gyr}{{\,{\rm Gyr}}}

\begin{document}

\title[Parametrising Star Formation Histories]{Parametrising Star Formation Histories}
\author[V.Simha et al]
{Vimal Simha$^{1}$, David H. Weinberg$^{2}$, Charlie Conroy$^{3}$, Romeel Dav\'{e}$^{4,5,6}$, 
\newauthor
Mark Fardal$^{7}$, Neal Katz$^{7}$, Benjamin D. Oppenheimer$^{8}$\\
$^1$ Institute for Computational Cosmology, Department of Physics,
Durham University, South Road, Durham DH1 3LE, United Kingdom\\
vimal.simha@durham.ac.uk\\
$^2$Astronomy Department and Center for Cosmology and AstroParticle Physics, Ohio State University, Columbus, OH 43210, USA\\
dhw@astronomy.ohio-state.edu\\
$^3$Department of Astronomy \& Astrophysics, University of California,
Santa Cruz, CA, USA\\
$^4$University of the Western Cape, Bellville, Cape Town 7535, South Africa\\
$^5$South African Astronomical Observatories, Observatory, Cape Town 7925, South Africa\\
$^6$African Institute for Mathematical Sciences, Muizenberg, Cape Town 7945, South Africa\\
rad@astro.as.arizona.edu\\
$^7$Astronomy Department, University of Massachusetts at Amherst, MA 01003, fardal,nsk@astro.umass.edu\\
$^8$Center for Astrophysics and Space Astronomy, Department of Astrophysical and Planetary Sciences,\\ University of Colorado,
389 UCB, Boulder, CO 80309, USA;
oppenheimer@strw.leidenuniv.nl\\
}

\maketitle


\begin{abstract}
We examine the star formation histories (SFHs) of galaxies in smoothed particle
hydrodynamics (SPH) simulations, compare them to parametric models that are
commonly used in fitting observed galaxy spectral energy distributions, and
examine the efficacy of these parametric models as practical tools for
recovering the physical parameters of galaxies.
The commonly used $\tau$-model, with
$\dot{M}_* \propto e^{-(t-t_i)/\tau}$, provides a poor match to the SFH of our
SPH galaxies, with a mismatch between early and late star formation that leads
to systematic errors in predicting colours and stellar mass-to-light ratios.
A one-parameter lin-exp model, with $\dot{M}_* \propto t\, e^{-(t-t_i)/\tau}$,
is much more successful on average, but it fails to match the late-time
behaviour
of the bluest, most actively star-forming galaxies and the passive, ``red and
dead'' galaxies.  We introduce a 4-parameter model, which transitions from
lin-exp to a linear ramp after a transition time $t_{\rm tr}$, which describes
our simulated galaxies very well. In practice, we can fix two parameters ($t_i$
and $t_{\rm tr}$) without significant loss of accuracy.  We test the ability of
these parametrised models to recover (at $z=0$, 0.5, and 1) the stellar
mass-to-light
ratios, specific star formation rates, and stellar population ages from the
galaxy colours, computed from the full SPH star formation histories using the
FSPS code of Conroy et al.\ (2009). Fits with $\tau$-models systematically
overestimate $M_*/L$ by $\sim 0.2$ dex, overestimate population ages by
$\sim 1-2$ Gyr, and underestimate $\dot{M}_*/M_*$ by $\sim 0.05$ dex. Fits with
lin-exp are less biased on average, but the 4-parameter model yields the best
results for the full range of galaxies. Marginalizing over the free parameters
of the 4-parameter model leads to slightly larger statistical errors than
1-parameter fits but essentially removes all systematic biases, so this is our
recommended procedure for fitting real galaxies.
\end{abstract}

\newpage
\section{Introduction}

Large samples of galaxies with multi-wavelength photometric data and
spectroscopic data \citep[e.g.][]{york00,gia04,scoville07} have
allowed galaxy evolution studies to shift from luminosity and colour
to the more physical plane of stellar mass and star formation rate (SFR),
and to examine
other aspects of galaxy evolution such as median ages of stellar
populations, importance of bursts, correlations with stellar and gas
phase metallicity, AGN activity, dust extinction etc. Deep photometric
surveys have also renewed emphasis on photometric redshifts, which
require a model of intrinsic galaxy colour. A crucial step in such
analyses is fitting
a parametric star formation history (SFH) to each galaxy's observed
spectral energy distribution (SED).
One of the most
commonly used parametrisations is the so called ``$\tau$-model,'' where the SFH
is described by an exponentially decreasing SFR
with e-folding time $\tau$
\citep[e.g.,][]{bruzual83,papovich01,shapley05,lee09,pozzetti10,foster09},
sometimes augmented with bursts
(e.g., \citealt{kauffmann03,brinchmann04}).
Some authors \citep[e.g.,][]{lee10} advocate an alternative model where the
SFH is parametrised by $t\,e^{-t/\tau}$, which allows linear growth
at early times followed by an exponential decline at late times. This
is sometimes referred to as the ``delayed" or ``extended"
$\tau$-model, but in this paper we shall refer to it as the lin-exp
(linear-exponential) model. Another approach is to fit the SFR in bins
of time (e.g., \citealt{panter07,tojeiro09}),
which has the virtue of generality but places strong demands
on the quality of the data and the accuracy of the population synthesis
models.

In this paper, we examine which parametrised models give good
descriptions for the SFHs of galaxies in smoothed particle hydrodynamics
(SPH) simulations. While SPH simulations are not a perfect
representation of the real Universe, they provide useful guidance on
what functional forms of the SFH may be necessary.
The simulations incorporate a wide range of processes
that may be important in galaxy SFHs, including accretion with environmental
dependence and stochastic variations, minor and major mergers, conversion of
gas to stars based on physical conditions in the interstellar medium,
ejection of gas in galactic winds, and recyclying of this ejected material
through subsequent accretion.

Our galaxy SFHs are obtained from a hydrodynamical simulation of a
cosmological volume ($50\hmpc$ cube, modeled with $2\times 288^3$ particles)
incorporating gas cooling, star formation, and galactic winds driven
by star formation.  The simulation
reproduces the observed stellar mass function
and HI mass function of galaxies quite well up to galaxies with stellar mass
$M_* \sim 10^{11} M_\odot$ \citep{oppenheimer10,dave13}, but it fails
at higher masses, predicting galaxies that are more massive than observed
and have too much late-time star formation.  To obtain a better match to the
observed galaxy stellar mass function and colour-magnitude
diagram, we also apply a post-processing prescription to the
simulation that has the
effect of quenching star formation in massive galaxies.
Throughout the paper we consider both the galaxy population predicted
directly by the simulation and the population that results from
applying this post-processing prescription; we refer to the former
as the ``Winds'' population and the latter as the ``Winds+Q'' population.

We compare parametric models for the SFH of galaxies to the SFH of
galaxies in our SPH simulation to investigate how well various parametric forms
describe the shape of the SFH of simulated galaxies and how well they
predict the colours and mass-to-light ratios of their stellar populations.
We then examine the effectiveness of these parametric models
as practical tools. We compute the colours of our simulated galaxies
using their true SFHs, then
fit parametric models to the colours and ask
how well these fits
recover physical parameters of interest such as the stellar mass,
population age, and current star formation rate. Our investigation offers
insight into the shortcomings of commonly used SFH models in regard to
the biases and errors they introduce in estimates of the physical
parameters of galaxies.
It also has practical import for future studies of galaxy
evolution, as we suggest a new parametric model
that describes the full variety of SFHs in our simulations, which
can be used straightforwardly to interpret observations of galaxy
populations.

Sources of systematic uncertainty in the SED fitting technique have
been investigated by several authors. For example, \cite{conroy09}
investigate uncertainties arising from the assumed form of the initial
mass function (IMF) and the treatment of stellar
evolution, \cite{papovich01} examine errors introduced by
uncertainties in the dust extinction, and \cite{lee09} find significant
errors and systematic biases when standard methods for inferring
ages and stellar masses of Lyman
Break Galaxies (LBGs) are applied to mock catalogues constructed
from semi-analytic models.
Obtaining accurate stellar spectral libraries that cover the full range of
stellar populations present in observed galaxies is a particular
challenge. (See \citealt{conroy13} for a review of the stellar population
synthesis technique, and inference of physical parameters of galaxies from
SEDs.) 
However, even if these sources of systematic
errors are controlled or eliminated, additional systematic
uncertainties arise from the assumed shape of the SFH,
and to get the most from the data one wants a model that has as much
flexibility as needed but not more than is needed.  It is this
aspect of population synthesis modeling that we focus on in this paper.

In \S2, we describe our
simulation, our method for identifying halos and galaxies,
our prescription for quenching star-formation in massive galaxies,
and our parametric SFH models.
In \S3, we fit these parametric models to the SFHs of our simulated
galaxies and ask how well they describe the SFHs
and the physical parameters that can be derived from
them. In \S4, we fit these models to the colours of galaxies and
compare the physical parameters obtained from these fits to their
``true'' values in the simulation.  We summarise our
results and discuss their implications in \S5.
The Appendix compares our parametric SFH model to the one
proposed recently by \cite{behroozi13}.

\section{Methods}

\subsection{Simulation}

Our simulation is performed using the GADGET-2 code \citep{springel05}
as modified by \cite{oppenheimer08}. Gravitational forces are
calculated using a combination of the Particle Mesh
algorithm \citep{hockney81} for large distances and the hierarchical
tree algorithm \citep{barnes86,hernquist87} for short distances. The
SPH algorithm is entropy and energy
conserving and is based on \cite{springel02}. The details of the
treatment of radiative cooling can be found in \cite{katz96} and
\cite{oppenheimer06}.
The details of the treatment of star formation can
be found in \cite{springel03}. Briefly, each gas particle satisfying a
temperature and density criterion is assigned a star formation rate,
but the conversion of gaseous material to stellar material proceeds
stochastically. The parameters for the star formation model are
selected so as to match the $z=0$ relation between star formation rate
and gas density \citep{kennicutt98,schmidt59}.

We adopt a $\Lambda$CDM cosmology (inflationary cold dark matter with
a cosmological constant) with $\Omega_m$=0.25,
$\Omega_{\Lambda}$=0.75, $h\equiv$ H$_0/$100 km
s$^{-1}$Mpc$^{-1}$=0.7, $\Omega_b=0.044$, spectral index $n_s=0.95$,
and the amplitude of the mass fluctuations scaled to
$\sigma_8=0.8$. These values are reasonably close to recent estimates
from the cosmic microwave background \citep{larson10} and large scale
structure \citep{reid10}. We do not expect minor changes in the values
of the cosmological parameters to affect our conclusions.

We follow the evolution of $288^3$ dark-matter particles and $288^3$
gas particles, i.e. just under 50 million particles in total, in a
comoving box that is $50\hmpc$ on each side, from $z=129$ to
$z=0$. The dark matter particle mass is 4.3 $\times$ $10^8$
$M_{\odot}$, and the SPH particle mass is 9.1 $\times$ $10^7$
$M_{\odot}$. The gravitational force softening is a comoving 5$h^{-1}$
kpc cubic spline, which is roughly equivalent to a Plummer force
softening of 3.5$h^{-1}$ kpc.
Higher resolution simulations of smaller volumes
(e.g., \citealt{dave13}) would yield more accurate SFH predictions
at a given stellar mass, but for the purposes of this paper we considered
it more important to have good statistics for a wide range of galaxy
masses and environments, so we chose to focus on a larger volume
simulated at lower resolution.  There are also uncertainties associated
with the hydrodynamics algorithm itself (e.g., \citealt{agertz07,sijacki12}),
but for our purposes these are less important than the physical
uncertainties associated with feedback and quenching mechanisms.
We discuss how all of these effects might impact our
conclusions in \S 5.

Our simulation incorporates kinetic feedback through momentum driven
winds as implemented by \cite{oppenheimer06,oppenheimer08},
where the details of the implementation can be found. Briefly, wind
velocity is proportional to the velocity dispersion of the galactic
halo, and the ratio of the gas ejection rate to the star formation
rate is inversely proportional to the velocity dispersion of the
galactic halo.  Except for the differences in volume and particle mass,
our simulation is similar to the ``vzw'' simulations
of \cite{oppenheimer10}, who investigate the growth of galaxies by
accretion and wind recycling and compare predicted mass functions to
observations. The specific simulation analyzed here was
also used by \cite{zu10} to
investigate intergalactic dust extinction and \cite{simha12} to
investigate subhalo abundance matching techniques.
The vzw model is quite successful at reproducing observations including
quasar metal absorption line statistics at high \cite{oppenheimer06}
and low \citep{oppenheimer12} redshift, the HI mass function
of galaxies at $z=0$ \citep{dave13}, and the galaxy stellar
mass function up to luminosities $L \sim L_*$ \citep{oppenheimer10}.

We identify dark matter haloes using a FOF (friends-of-friends)
algorithm
\citep{davis85}. The algorithm selects groups of particles in which
each particle has at least one neighbour within a linking length, set to the interparticle separation at one-third of the virial overdensity , which is calculated for the value of $\Omega_M$ at each redshift using the fitting formula of \cite{k96}. 
Many of our plots
distinguish between the behavior of central galaxies of halos and
satellite galaxies (see \citealt{simha09} for discussion).
The most massive object in a FOF halo is referred to as a central galaxy
and the others as satellites.

Hydrodynamic cosmological simulations that incorporate cooling and
star formation produce dense groups of baryons with sizes and masses
comparable to the luminous regions of observed
galaxies \citep{katz92,evrard94}. We identify galaxies using the
Spline Kernel Interpolative DENMAX
(SKID\footnote{http://www-hpcc.astro.washington.edu/tools/skid.html})
algorithm \citep{gelb94,katz96}, which identifies gravitationally
bound particles associated with a common density maximum. We refer to
the groups of stars and cold gas thus identified as galaxies. The
simulated galaxy population becomes substantially incomplete below a
threshold of $\sim$64 SPH particles \citep{murali02}, which
corresponds to a baryonic mass of 5.8 $\times 10^9$ $M_{\odot}$. For
this work, we adopt a higher stellar mass threshold of $10^{10}$
$M_{\odot}$ because star formation histories of lower mass galaxies
are noisy, even if their final stellar masses are reasonably robust.

For each SKID identified galaxy at $z=0$, we trace the formation time
of its stars. We then bin these star formation events in time to
extract a star formation rate as a function of time. From this SFR$(t)$,
we
generate colours using the stellar population synthesis package
FSPS \citep{conroy09}. We assume solar metallicity and
ignore dust extinction in this paper. While dust exinction and metallicity
should be additional free parameters when fitting SFHs to colours of
observed galaxies, our goal in this paper is to isolate
the impact of SFH shape, so we avoid the step of inserting and
then attempting to remove these additional effects.

\subsection{Quenching Model}

\begin{figure}
\centerline{
\epsfxsize=5.5truein
\epsfbox{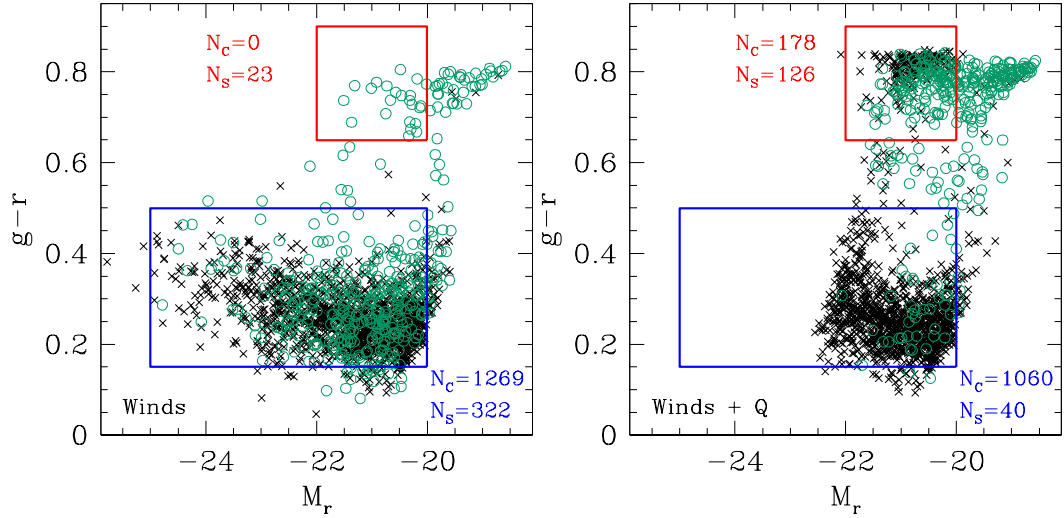}
}
\caption{
The colour-magnitude diagram of galaxies in our SPH simulation (Winds; left)
and after applying our post-processing, quenching prescription (Winds +Q;
right).
Each point is an individual galaxy. Green open circles are satellite
galaxies and black crosses are central galaxies.
Since saturation makes it difficult to judge the relative numbers of
central and satellite galaxies, we list these numbers for the two
boxed regions in each panel.  Note in particular that the red sequence
is almost entirely populated by satellite galaxies in the
Winds model but is dominated (at the bright end) by central galaxies
in the Winds+Q model. $N_C$ and $N_S$ denote the number of central and
satellite galaxies in each box, respectively.
}
\label{fig:sub1}
\end{figure}

The left panel of Figure \ref{fig:sub1} shows the $(g-r)$ colour of
SPH galaxies against their $r$-band magnitudes. Each point is an
individual galaxy. Central galaxies are shown as black crosses and
satellite galaxies as green open circles. The red galaxies in our
simulation are almost all low luminosity satellites. In common with
other hydrodynamic simulations that do not include AGN feedback, our
simulation fails to produce bright red galaxies. Our simulation also
matches the observed galaxy luminosity function up to $L_*$ but
overpredicts the number density of galaxies brighter than $L_*$. To
obtain bright red galaxies and match the observed stellar
luminosity function, we construct a ``quenched winds'' (Winds + Q)
population by implementing a post-processing prescription.

Various lines of observational evidence suggest that star formation is
quenched in high mass halos. The most commonly invoked explanation is
AGN feedback. This could be connected to the transition between cold
and hot mode gas accretion \citep{keres05,dekel06},
with AGN feedback being more effective in
suppressing the accretion of the hot gas that appears in higher mass halos.
In any case, models of
galaxy formation that match the observed luminosity function or
stellar mass function
have some quenching mechanism for central galaxies of high mass halos.

In our simulation we have a complete history of star formation events
for each galaxy. We modify the SFR of galaxies based on their parent
halo mass at the epoch of each star formation event. In halos more
massive than a halo mass threshold, $M_{\rm max}$, we eliminate all
star formation events.  In less massive halos we multiply the mass
of stars formed by a factor that scales linearly with halo mass down
to a halo mass $M_{\rm min}$, below which we do not alter the star
formation rate.  The effect of this post-processing can be described as
\begin{equation}
\begin{rm} SFR (winds + Q) = SFR (winds) \end{rm} \times f(M_H)
\end{equation}
where $M_H$ is the parent halo mass and
\begin{equation}
f(M_H) =
\begin{cases}
1 & \mbox{if } (M_H < M_{\rm min})\\ 
{{\Delta M}/({M_{\rm
max}-M_{\rm min}})} & \mbox{if }(M_{\rm min} < M_H < M_{\rm max})\\ 
0
& \mbox{if } (M_H > M_{\rm max})~
\end{cases}
\end{equation}
and
\begin{equation}
\Delta M = M_{\rm max}-M_H~.
\end{equation}
We implement this procedure for both central and satellite galaxies.

\begin{figure}
\centerline{
\epsfxsize=5.5truein
\epsfbox{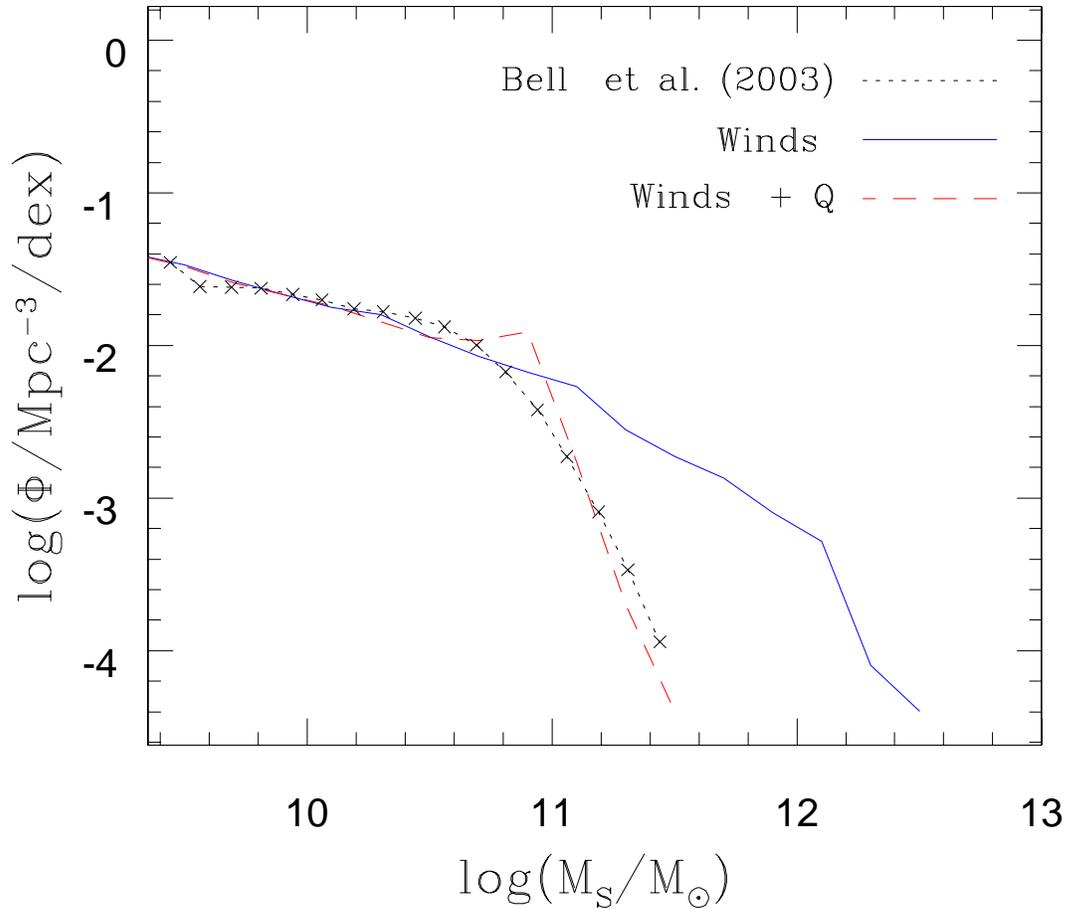} 
}
\caption{
The galaxy stellar mass function at $z=0$ in our simulation (solid), and
after applying our quenched winds post-processing prescription (dashed)
compared to the observations of \protect\citealt{bell03} (dotted). In this and later
plots, M$_S$ refers to the stellar mass of SKID-identified galaxies.
}
\label{fig:sub2}
\end{figure}

We set $M_{\rm min}$ equal to 1.5 $\times 10^{12}$ $M_{\odot}$ and
$M_{\rm max}$ equal to 3.5 $\times 10^{12}$ $M_{\odot}$. These
parameters are chosen to obtain an approximate match to the observed
stellar mass function. \cite{oppenheimer10} discuss the comparison
between the predicted stellar mass functions and observational
estimates in some detail. Roughly speaking, our simulation reproduces
observational estimates for $M_*$ $<$ 10$^{11}$ M$_{\odot}$ , but it
predicts excessive galaxy masses (at a given space density) for
$M_* > 10^{10.8} M_{\odot}$. Figure \ref{fig:sub2} shows the galaxy
stellar mass function for the simulation (Winds) and for the
post-processing prescription implemented here (Winds + Q). While we do
not obtain a perfect match, Winds+Q is a substantial improvement over
the original simulation results.

The right panel of Figure \ref{fig:sub1} shows the colour-magnitude
diagram for the quenched
winds population. While our post-processing does
produce a ``massive red sequence'' of central galaxies,
we do not match the detailed
properties of the observed colour-magnitude diagram such as its
slope. Also, the brightest galaxies in our model are blue, while the
brightest observed galaxies are red.
(The most massive galaxies in Winds + Q are on the red sequence,
but their higher mass-to-light ratios make them less luminous than
the most massive blue galaxies.)
\cite{gabor10} perform a
detailed investigation of various post-processing prescriptions,
including one similar to our own, and compare them to the observed stellar
mass function, slope of the colour-magnitude diagram and other
observables. We refer the reader to that paper for a more in depth
discussion of physical and observational issues. Clearly our Winds + Q
model is not perfect, but it serves our purpose of providing a model
population of galaxies that is a reasonable match to observations,
including a large population of quenched massive galaxies as in the
real universe.

\subsection{SFH models}

The simplest SFH model we consider is the ``$\tau$-model'' where the
SFH is described by an exponentially decreasing function with
timescale $\tau$, starting at time $t_i$:
\begin{equation}
{\rm SFR}(t-t_i) = Ae^{-(t-t_i)/\tau}.
\end{equation}
Our
simulated galaxies show little star formation before 1 Gyr, and
we therefore set $t_i=1\Gyr$; choosing $t_i=0$ would make the fits to
the simulated SFHs systematically worse.

In practice the simulated SFHs rise to a maximum rather than starting
at a high value, so we also consider the lin-exp model with
\begin{equation}
{\rm SFR}(t) = A(t-t_i)e^{-(t-t_i)/\tau} .
\label{eq:lin-exp}
\end{equation}
As with the exponential model, we treat $\tau$ as a free parameter
and set $t_i=1\Gyr$.
In the limit of $\tau \gg t_0$, the lin-exp SFH is simply a
linearly rising SFR$(t)$, while in the limit $\tau\rightarrow 0$
it is a burst at $t=t_i$.

For greater generality
we consider a lin-exp model at early times
that transitions to a linear ramp at late times:
\begin{equation}
{\rm SFR}(t) =\begin{cases}
{A(t-t_i)e^{-(t-t_i)/\tau}} & \mbox{for } (t \leq t_{\rm trans})\\
{{\rm SFR}(t_{\rm trans}) + {\Gamma(t-t_{\rm trans})}}
& \mbox{for } (t > t_{\rm trans}) .\\
\end{cases}
\label{eq:general}
\end{equation}
The key feature of this model is that it decouples the late-time SFR
(after $t_{\rm trans}$) from the early-time SFR, though it requires
continuity at $t_{\rm trans}$.  The new parameter $\Gamma$ determines
the slope of the SFH at $t>t_{\rm trans}$, allowing rising, flat, or
falling SFR$(t)$.
(In the FSPS code, $\Gamma$ is referred to as ``$\tan\theta$'',
where $\theta$ is the angle of the linear ramp in the SFR-$t$ plane.)
We set SFR$=0$ at times
when eq.~\ref{eq:general} gives a negative result, thus permitting
a truly truncated SFH.
We can describe the SFHs of our simulated galaxies adequately
by setting $t_i=1\Gyr$ and $t_{\rm trans}=t_0-3.5\Gyr = 10.7\Gyr$, so that
$\Gamma$ is the only new free parameter.  We refer to this as
our ``2-parameter model.''  Modest changes in $t_{\rm trans}$ would lead
to changes in $\Gamma$ values but would not significantly degrade
the fits.  However, the timing of the SFR transition and the onset of early
star formation could be affected by the specific feedback physics and
numerical resolution in our simulations, so real galaxies may have greater
variety.  We therefore also consider a 3-parameter model in which
$t_{\rm trans}$ is a fitting parameter and a 4-parameter model in which
both $t_{\rm trans}$ and $t_i$ are fitting parameters.  The 4-parameter
model is the most general one we consider in this paper and the one
we advocate for practical applications.  Note that lin-exp is a special
case of the 4-parameter model with $t_{\rm trans}=t_0$ and
$t_i=1\Gyr$.

During the course of our investigation, we also explored other
possibilities. For example, we considered models like
$t^{\alpha}e^{-t/\tau}$, but rejected them because they added more
complexity without significantly improving the performance in describing
the SFHs of our simulated galaxies.
We also considered other simple extensions of the lin-exp model
such as adding a constant late-time component instead of a linear ramp.
However, this model proved
insufficient to describe the SFH of galaxies in our simulation,
which sometimes show a truncation or a rising late-time SFR.
Because our simulations rarely show discrete ``bursts'' of star
formation, we did not investigate parametrisations like the
$\tau$+burst models of \cite{kauffmann03}.

\section{Fitting Models to Simulated SFHs}

In this Section, we fit the five parametric models described in \S2.3, namely,
the $\tau$-model, lin-exp model, and the 2, 3, and 4-parameter models to the
SFHs of galaxies in the SPH simulation (the ``Winds'' population) and
the post-processed simulation (the ``Winds+Q'' population).
We choose model parameters to minimize the cost function
\begin{equation}
C = \int_0^{t_{\rm max}} \sqrt{(\rm{SFR}-{\rm{SFR}}_{model})^2}~dt.
\label{eqm}
\end{equation}
We also impose an integral constraint:
\begin{equation}
\int_0^{t_{\rm max}} {{\rm SFR_{model}}}(t)~dt =
\int_0^{t_{\rm max}} {{\rm SFR}}(t)~dt~.
\end{equation}
The mass-to-light ratio, age quantiles of the stellar population, and predicted
colours are independent of the normalization of the SFH and are fully
determined by the shape alone.

\begin{figure}
\centerline{
\epsfxsize=5.0truein
\epsfbox{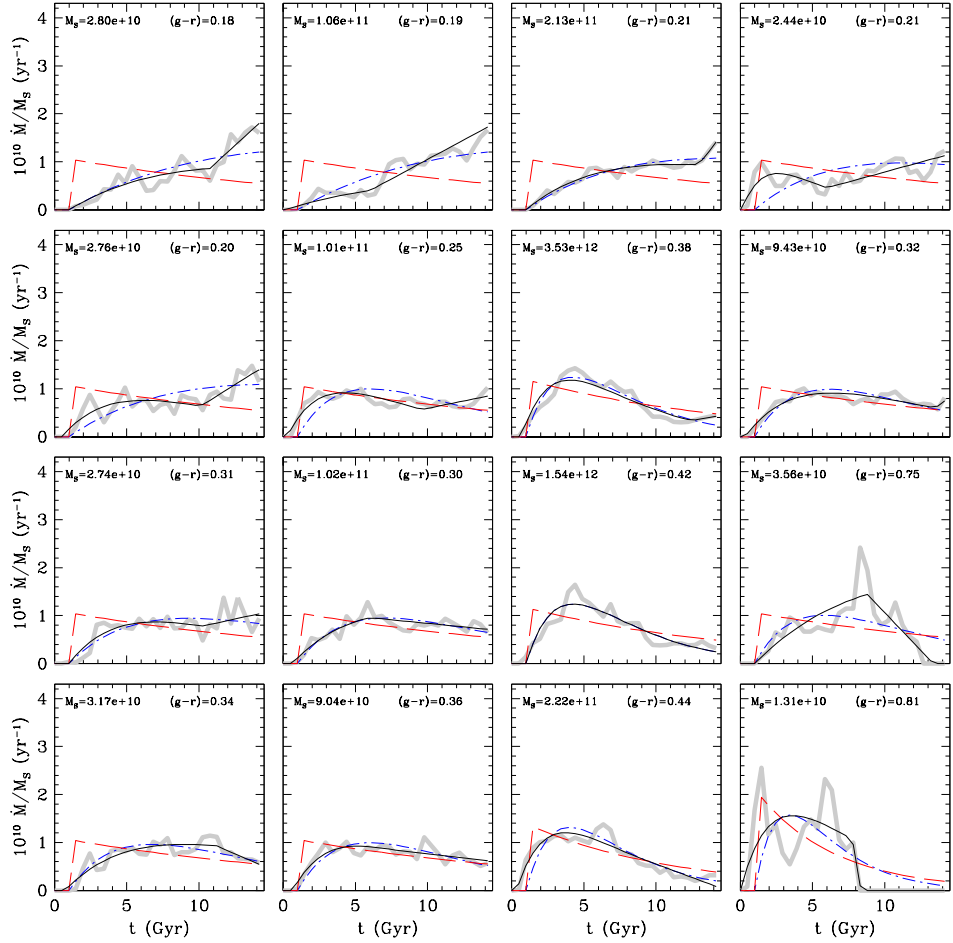} 
}
\caption{
Galaxy SFR versus time in the winds simulation. Each panel shows an
individual galaxy. The thick gray curve shows the SFR in the
simulation, the black solid curve shows the best-fit 4-parameter model
(see text), the blue curve shows the best-fit lin-exp model, and the
red curve shows the $\tau$-model. The first three columns from the
left show central galaxies in three mass bins, with mass increasing
from left to right, and the right column shows satellite galaxies. The
top row shows blue galaxies and each successive lower row shows a
redder colour.  }
\label{fig:sub3.1}
\end{figure}

Figure \ref{fig:sub3.1} shows the SFH of a representative selection of
SPH galaxies. SFR normalized by the stellar mass at $z=0$ is shown on
the vertical axis against time on the horizontal axis. The thick gray
solid curve shows the SFH in the simulation, and the other curves show
the best-fit models of the different SFH parametrisations.
The top row shows blue galaxies, and each
successive lower row shows a redder colour. The first three columns
from the left show central galaxies in three mass bins, with mass
increasing from left to right. The right column shows satellite
galaxies, where a satellite galaxy is defined as a SKID group that is
not the most massive galaxy in its FOF halo.

An examination of Figure \ref{fig:sub3.1} reveals several interesting
trends. The SFHs of low mass galaxies show bumps and wiggles,
but for the most part
the SFHs of individual galaxies are smooth, not
punctuated by starbursts and gaps.
Other implementations of star formation and feedback physics might
lead to burstier behaviour, but if star formation and its associated
outflows largely keep pace with accretion as they do in this
simulation, then a smooth SFH is the generic outcome.
At high $z$, most simulated galaxies have a
gradually increasing SFH, in contrast to  the steep increase
followed by exponential decline that is mandated by the
$\tau$-model. The shape of the peak in the SFH is generally matched by
the lin-exp model when we allow a start time of 1 Gyr for star
formation to commence.
While the slope of SFR$(t)$ at high redshift varies strongly from
galaxy to galaxy, there
is little variation around $t_i \approx 1\Gyr$; in particular, we find
no examples of galaxies that wait several Gyr before starting to
form stars.
At low $z$, the blue galaxies in the top two
rows have a rising SFR, while the red galaxies in the lower rows have
a falling SFR.  The lin-exp model often describes these histories
fairly well, but in some cases it cannot, such as the top two panels
in the left column and the bottom two panels in the right column.

As shown by \cite{simha09}, many satellite galaxies in these simulations
continue to accrete gas and form stars, in agreement with inferences from
observations \citep{weinmann06,weinmann10,wetzel13}.
The top two panels in the right column of
Figure \ref{fig:sub3.1} show two such examples. The third row of the
rightmost column shows a satellite galaxy that is not forming stars at
$z=0$. The bottom row of the same column shows a more extreme example,
where the SFR is truncated at $t$ $\sim$ 8 Gyr, after the galaxy falls
into a massive halo. The lin-exp model fails to match these truncated
SFHs, predicting a SFR that is too high at $z=0$ and consequently a
colour that is too blue compared to the simulation.
For low mass galaxies in the Winds+Q model, we find similar trends
to those in Figure~\ref{fig:sub3.1}, but for higher mass
galaxies the SFR at late times is systematically
lower, and for the most massive galaxies it is truncated before $z=0$.

\begin{figure}
\centerline{
\epsfxsize=5.0truein
\epsfbox{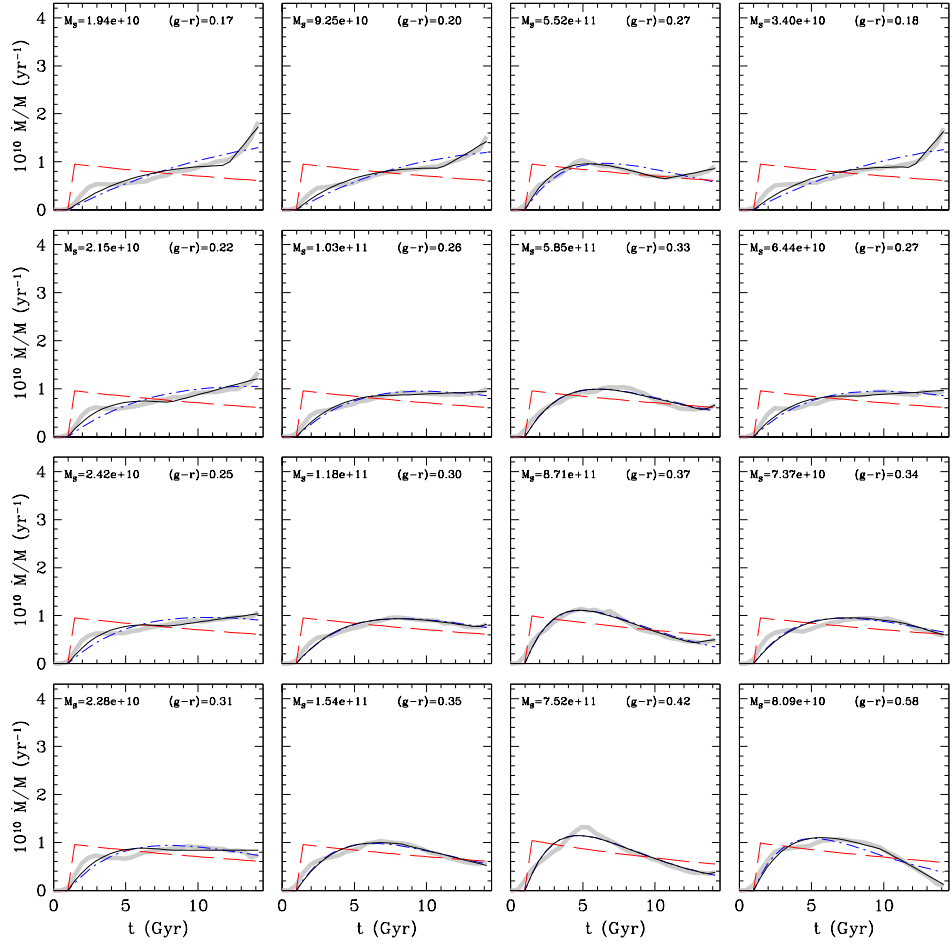} 
}
\caption{
Like Figure 3, but now showing the average SFH of central galaxies in
bins of mass and colour (left three columns) and of satellite galaxies
in bins of colour (right column).
The three central galaxy mass bins are chosen to contain approximately
equal numbers of galaxies, and in each column the four panels show the
four quartiles of the colour distribution in that bin.  Labels indicate
the mean stellar mass in each bin and the colour computed by FSPS
from the average SFH.
Note that the dot-dashed curve
(lin-exp) is sometimes fully obscured by the solid curve (4-parameter
model), and that the 4-parameter model fit is itself often obscured
by the true average SFH.}
\label{fig:sub3.2}
\end{figure}

Figure \ref{fig:sub3.2} shows the average SFH of galaxies in bins of
mass and colour chosen to contain approximately equal numbers of
galaxies.  As in Figure~\ref{fig:sub3.1}, the first three columns
show central galaxies ordered by increasing mass, the fourth
column shows satellite galaxies, and the rows are ordered from
the bluest quartile to the reddest quartile in each bin.
We also show the average stellar mass and the $g-r$ colour obtained
by treating the average SFH as the SFH of an individual galaxy and
using it as input for the stellar population synthesis code.
These curves are smoother than those in Figure~\ref{fig:sub3.1}
because they average over variations in individual SFHs, but they
reveal the same trends.  The $\tau$-model shows the same
systematic failures seen in in Figure~\ref{fig:sub3.1}.
The lin-exp model gives a good description of the average SFH
in most bins, but it underpredicts the $z=0$ SFR in the bluest galaxies
and overpredicts the $z=0$ SFR in red satellites.
These discrepancies lead to systematic deviations in the predicted colours
as shown below.

Figure \ref{fig:sub3.3} is similar to Figure \ref{fig:sub3.2} but
for the Winds + Q model. Results for the lowest mass central galaxies
are similar, of course, but for more massive galaxies the SFHs are
often truncated at late times and correspondingly more sharply peaked
at early times.  The lin-exp model is remarkably successful at
describing the SFH shape in most of these bins, capturing the
correlation between rapid early growth and suppressed late-time star
formation.  However, it fails to predict the correct $z=0$ SFR in some
cases.  We have shown results from the 4-parameter model in
Figures~\ref{fig:sub3.1}-\ref{fig:sub3.3}, but the results are only
slightly degraded if we fix $t_i=1\Gyr$ (3-parameter model) and
$t_{\rm trans}=10.7\Gyr$ (2-parameter model).

\begin{figure}
\centerline{
\epsfxsize=4.5truein
\epsfbox{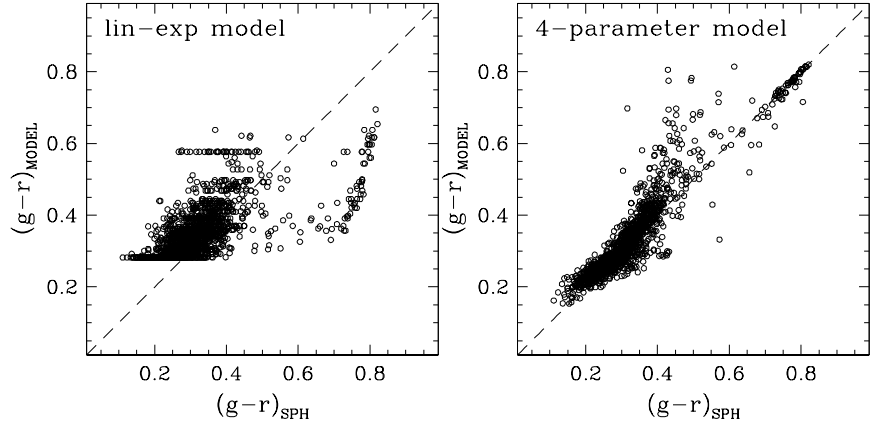} 
}
\caption{
(Left) Colour predicted by the best-fit lin-exp model versus the colour
computed from the full SFH of the corresponding SPH galaxy, at $z=0$.
(Right) Same, for the best-fit 4-parameter model.
Both panels are for the Winds galaxy population.
There are 1,828 individual galaxies in the plot.
The horizontal ridges in the left panel correspond to fits with
very long or very short timescale $\tau$.
}
\label{fig:sub3.4}
\end{figure}

The left panel of Figure \ref{fig:sub3.4} compares the $g-r$ colour
predicted by the best-fit lin-exp model to the SPH $g-r$ colour. While
computing the colours, we ignore
dust extinction and use the same SSPs to compute colours from the SFHs for
the SPH galaxy and the model fits, and therefore ignore the possibility of
template mis-match. Each
point is an individual galaxy. Because lin-exp is unable to match
the late time increase in SFR for the bluest galaxies, it predicts colours
that are systematically too red when $(g-r)_{\rm SPH} \leq 0.3$.
Conversely, for galaxies that are very red,
particularly satellite galaxies, it fails to match the truncation in
the SFH, instead predicting ongoing star formation and hence colours
that are too blue.
This error is particularly noticeable for galaxies with
$(g-r)_{\rm SPH} \geq 0.6$.
For comparison, the colour from the best-fit
4-parameter model is shown in the right panel. The late time linear
component with variable slope helps overcome both these shortcomings
of the lin-exp model, yielding accurate colour predictions for
the bluest and reddest galaxies.  For $(g-r)_{\rm SPH} = 0.45-0.6$ the
model colours are still systematically too red.
(Recall that all colours in the paper are computed for zero dust reddening.)

\begin{figure}
\centerline{
\epsfxsize=5.0truein
\epsfbox{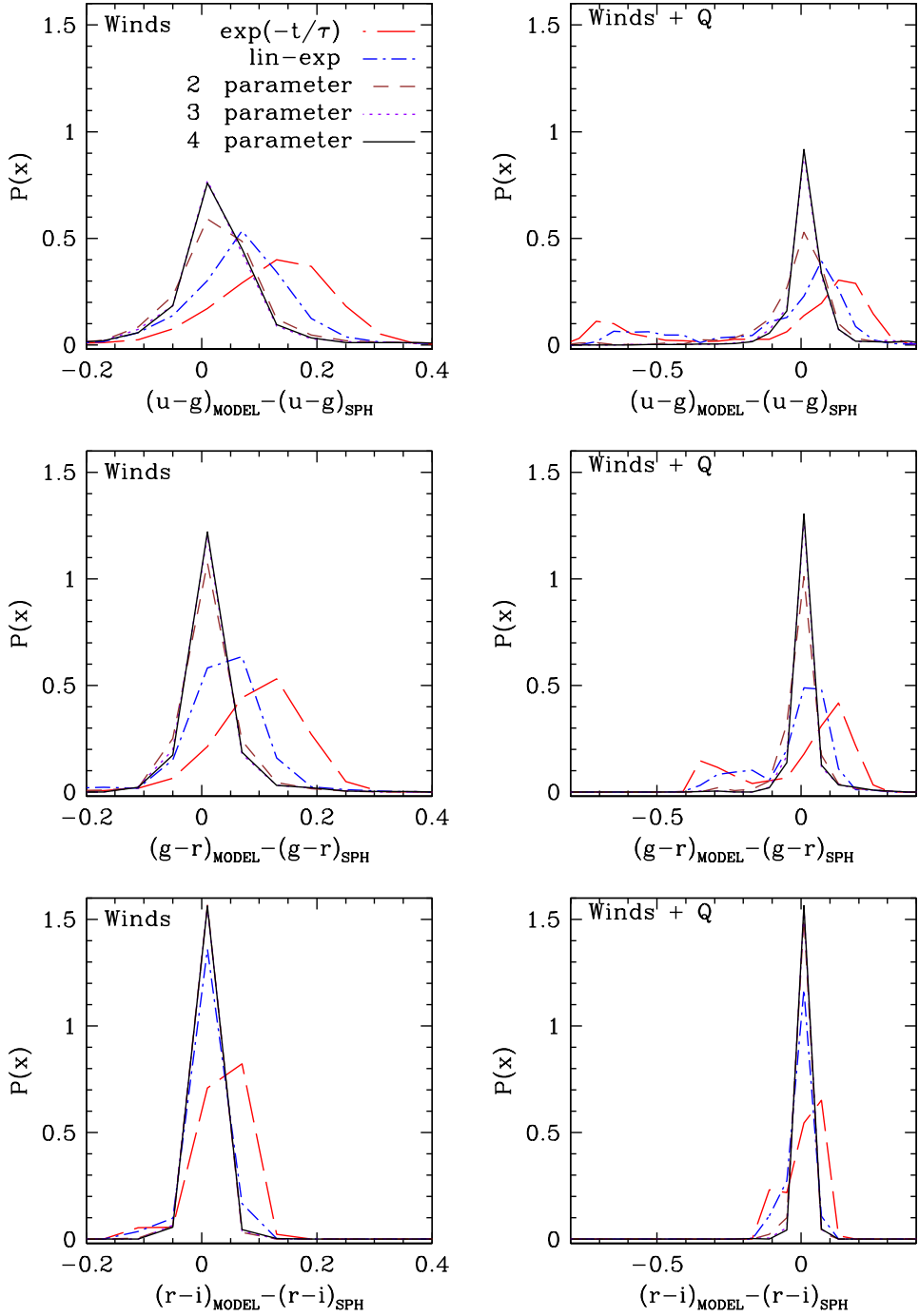} 
}
\caption{
Distribution of differences between parametric model colour and SPH
galaxy colour in the winds simulation (left) and the Winds + Q model
(right), at $z=0$.
Each curve stands for a different parametric model, and the
curves are normalised to unit integral. We ignore the effects of dust
extinction on the colours.
}
\label{fig:sub3.7}
\end{figure}

Figure \ref{fig:sub3.7} shows the distribution of the differences between
the colour of the best-fit parametric model and the SPH galaxy whose
SFH is fit. We show the 2-parameter and 3-parameter models in addition to the
three models shown in Figures~\ref{fig:sub3.1}-\ref{fig:sub3.3}.
The $\tau$-model requires too much early star formation relative
to late star formation and, therefore, predicts colours that are
systematically too red, by $\sim 0.15$ magnitudes in $u-g$,
$\sim 0.12$ magnitudes in $g-r$, and $\sim 0.05$ magnitudes in $r-i$.
The lin-exp model is mildly biased
towards redder colours, but a considerable improvement on the
$\tau$-model.
Results for the 2, 3, and 4-parameter models are nearly identical
and sharply peaked around the colour predicted using the galaxy's
true SFH.  For the Winds+Q population (right hand panels) the $\tau$
and lin-exp models have a low amplitude tail of galaxies whose
model colours are much too blue; these are the galaxies with truncated
SFHs.  This tail is strongly suppressed in the multi-parameter models.
As expected, colours at redder wavelengths
are predicted more accurately in every case
because they are less sensitive to late-time star formation.

\begin{figure}
\centerline{
\epsfxsize=4.0truein
\epsfbox{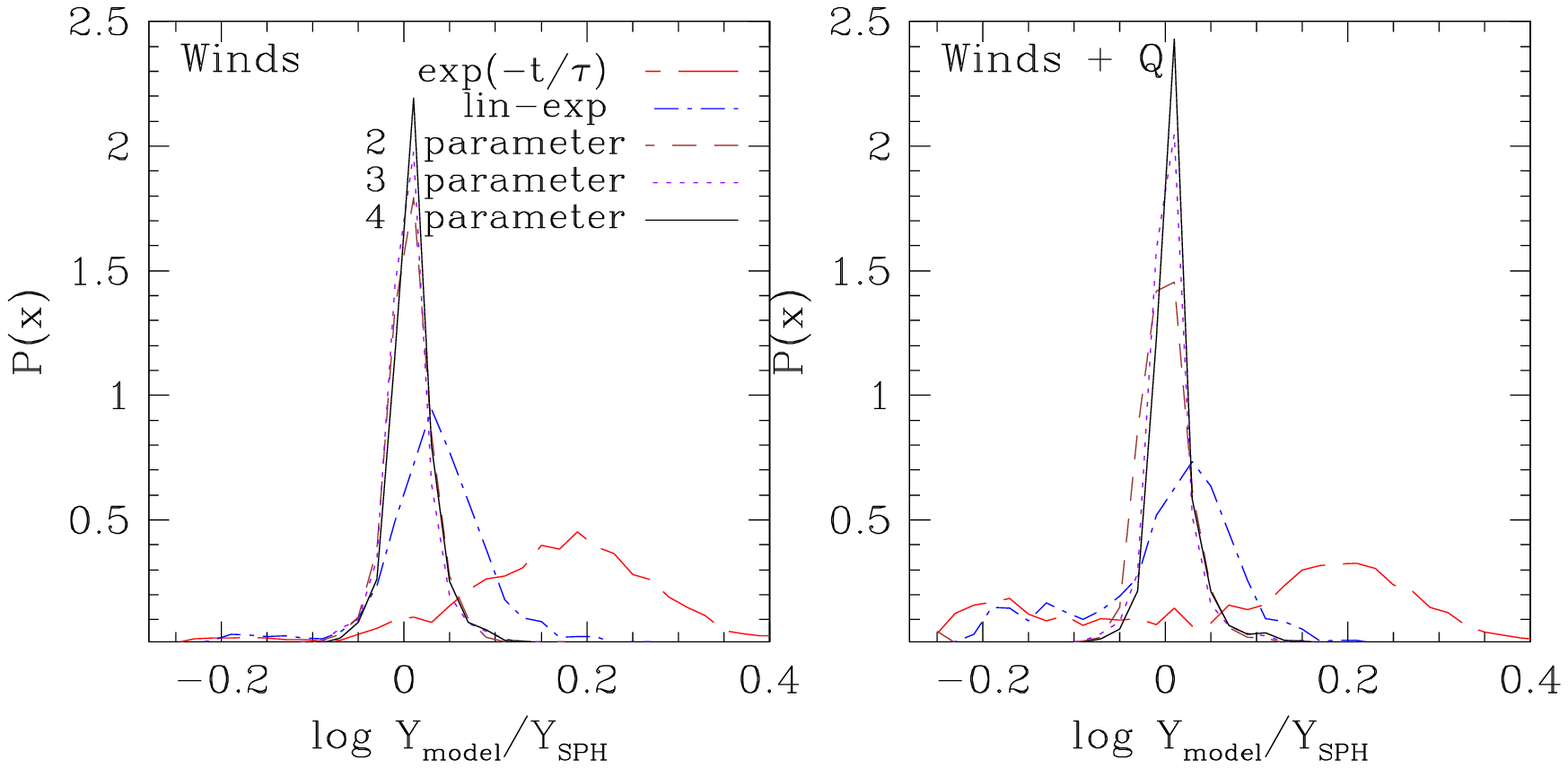} 
}
\caption{
Distribution of differences between the $r$-band mass-to-light ratio
predicted by the parametric models and that of the corresponding SPH
galaxy, at $z=0$.  Each curve stands for a different parametric model, and the
curves are normalised to unit integral.
}
\label{fig:sub3.9}
\end{figure}

One of the most important applications of SED fitting is to infer
the mass-to-light ratios of stellar populations, so that observed
luminosities can be converted to stellar masses.
Figure \ref{fig:sub3.9} compares the $r$-band stellar mass-to-light ratio
$Y_r \equiv M_*/L_r$  of
SPH galaxies to that obtained from various parametric fits to the SFH.
Specifically, we use FSPS to compute $r$-band luminosities from
either the simulated SFH or the SFH of the best-fit
parametric model (which is always constrained to reproduce the
simulated galaxy's $M_*$).
Because the best-fit $\tau$-model consistently has too much
early-time star formation and too little late-time star formation
(Figs.~\ref{fig:sub3.1} and~\ref{fig:sub3.2}), the $\tau$-model
fits systematically overestimate $Y_r$, with a typical offset
of 0.2 dex ($Y_{\rm model}/Y_{\rm SPH} \sim 1.6$).
The lin-exp model fares much better, producing a
reasonable match to the
mass-to-light ratio of most galaxies but overestimating $Y_r$ for
blue galaxies that have an increasing SFR at late times.
The 2, 3 and 4-parameter models
yield mass-to-light ratios sharply peaked around the true values, fitting
68\% of galaxies to within 6\%.  The right panel shows results
for the Winds+Q galaxy population.
In addition to the previous trends, the $\tau$ and lin-exp models
now have a tail of galaxies for which $Y_r$ is underestimated by up to 0.2 dex.
These are the red galaxies with sharply truncated SFH, which are poorly
represented in these models.  The 2, 3, and 4-parameter models, on the
other hand, can all represent these truncated SFHs, so they
do not produce a tail of underestimated mass-to-light ratios.


We examine the SFHs of galaxies at $z$ $\ge$ 0 to investigate
whether the parametric models that give good descriptions of $z=0$ SFHs
also do so at $z=0.5$ and
$z=1$. We fit the five parametric models described in \S2.3 to the
SFHs of galaxies in the SPH simulation at $z=1$ and $z=0.5$. In
addition to the lin-exp model described in eq.~\ref{eq:lin-exp} and
the general 4-parameter model described in eq.~\ref{eq:general},
we also fit a 3-parameter model where we fix $t_i$ = 1 Gyr and
a 2-parameter model where, in addition, we scale
$\ttr$ with the age of the Universe.
Specifically, we set
\begin{equation}
\ttr (z=z') = \ttr(z=0)~ \times ~t_{z'}/t_0
\label{eq:tscale}
\end{equation}
where $t_0$ is the age of the Universe at $z=0$,
$\ttr(z=0)=10.7$ Gyr is the value previously adopted in our
2-parameter model,
and $t_{z'}$ is the age of the Universe at redshift $z=z'$.
The 2-parameter model SFH follows a lin-exp model for the
first 75\% of its lifetime, and a linear ramp thereafter.
We restrict our analysis to the Winds + Q model.

\begin{figure}
\centerline{
\epsfxsize=5.0truein
\epsfbox{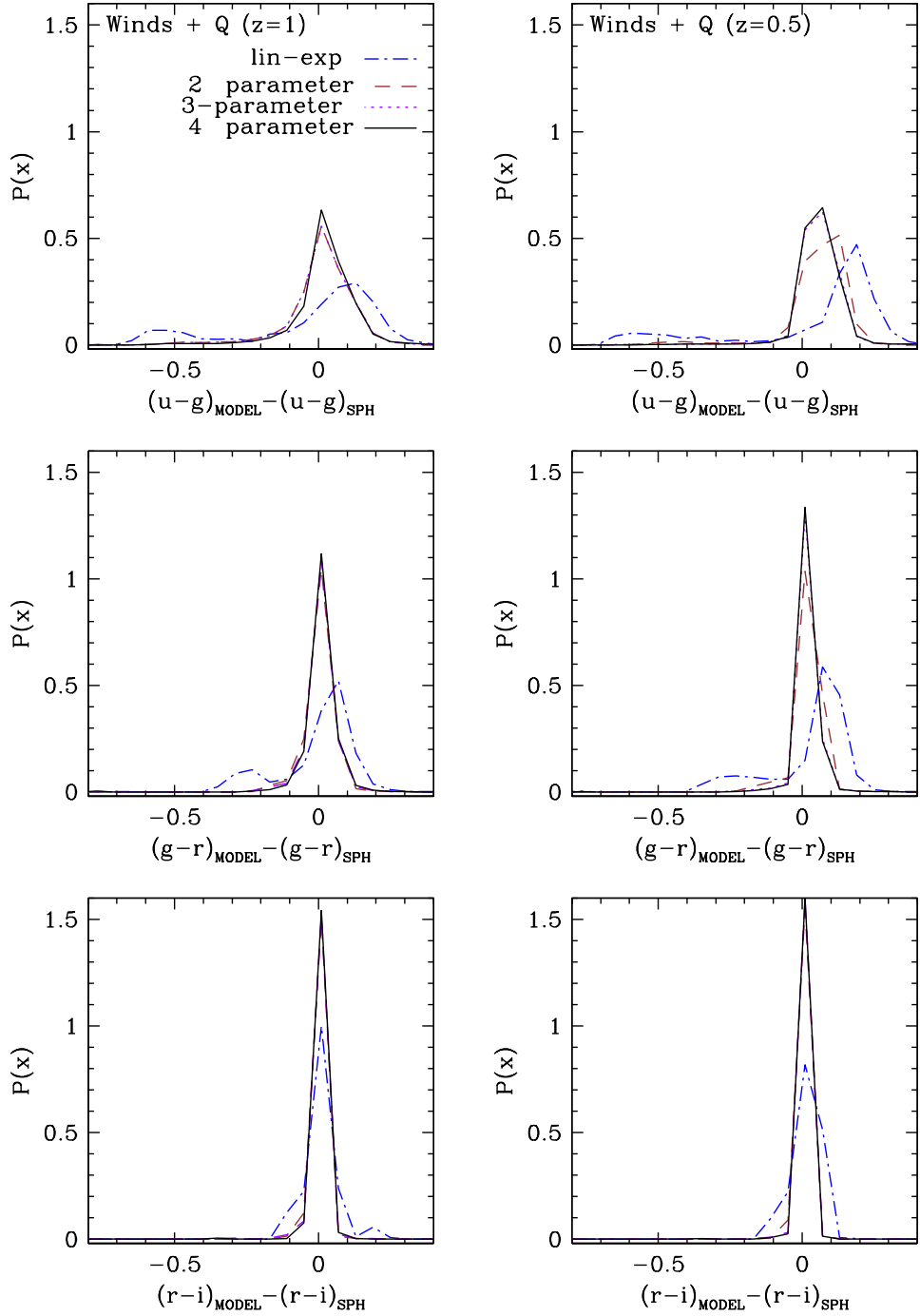} 
}
\caption{
Distribution of differences between parametric model colour and SPH
galaxy colour at $z=1$ (left) and at $z=0.5$
(right). Each curve stands for a different parametric model, and the
curves are normalised to unit integral.
}
\label{fig:sub3.10}
\end{figure}

Figure \ref{fig:sub3.10} shows the distribution of differences between the
colour of the
best-fit parametric model and the SPH galaxy whose SFH is fit at $z=1$ (left)
and $z=0.5$ (right) for the Winds + Q model.  The lin-exp model is generally
biased towards redder colours with the exception of a small number of galaxies
with truncated SFHs whose model colours are too blue. The 4-parameter model is
sharply peaked around the colour predicted from the galaxy's true SFH. Fixing
$t_i$ = 1 Gyr (3-parameter model) produces a nearly identical fit as it is
close to the value of $t_i$ obtained from the best-fit 4-parameter model
for most galaxies, and in any case small changes in the SFR at early times
do not have an appreciable effect on the colour. 
Further restricting to the 2-parameter model
only marginally degrades the fits to the SFH. 
While the value of $\ttr$ fixed according to eq.~\ref{eq:tscale} differs
slightly 
from the mean $\ttr$ of the best-fit 4-parameter model to all galaxies,
correlations between $\ttr$ and $\Gamma$ ensure that the best-fit 4-parameter
SFH and the best-fit 2-parameter SFH are similar even when they have a
different $\ttr$. All models predict the redder colours more accurately
because they are less sensitive to late-time star formation. As expected,
the $\tau$-model (not shown) performs worse at high redshift than at $z=0$,
predicting colours that are too red because it requires too much early star
formation relative to late star formation.

\begin{figure}
\centerline{
\epsfxsize=4.0truein
\epsfbox{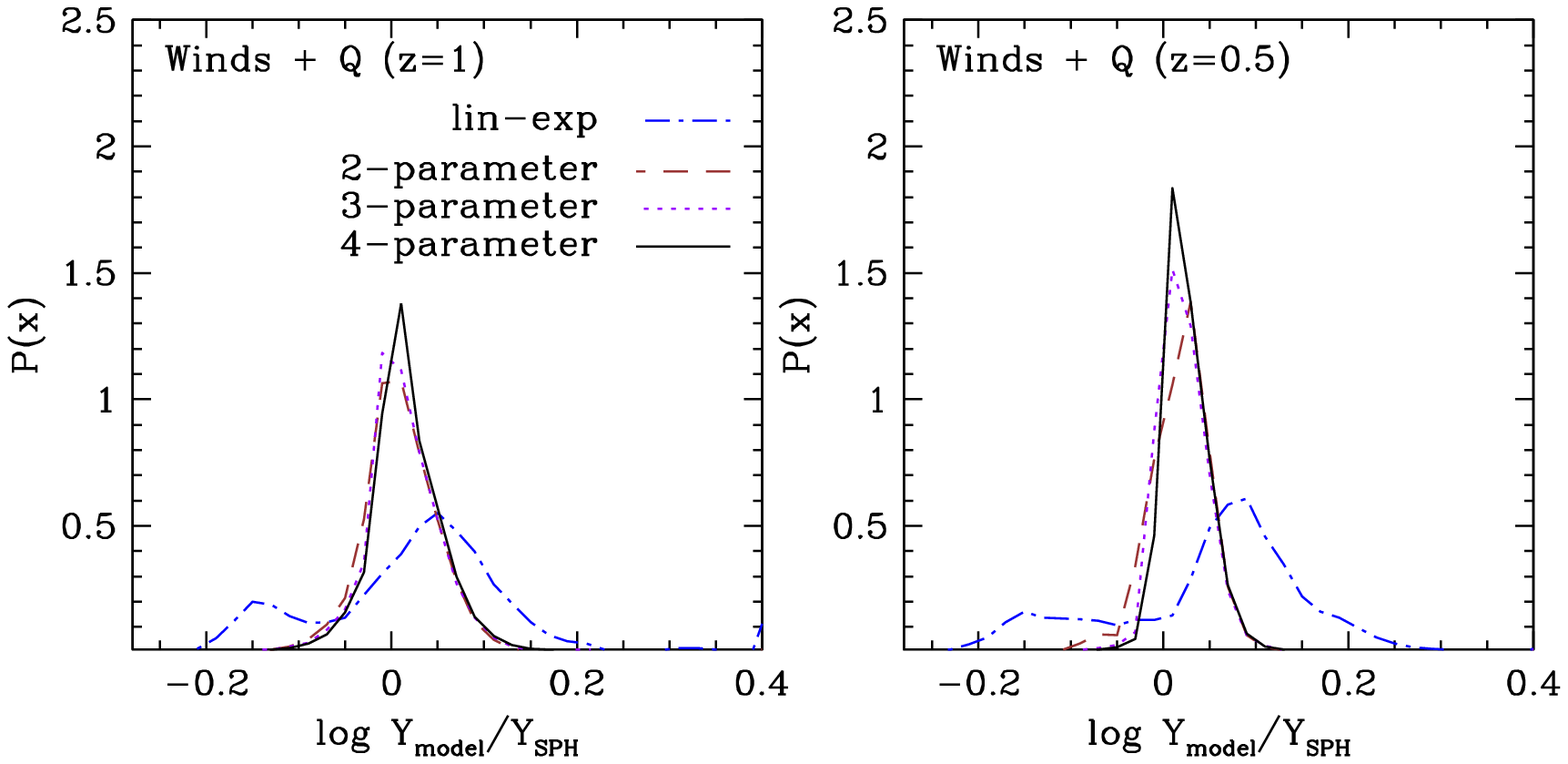} 
}
\caption{
Distribution of differences between the $r$-band mass-to-light ratio
predicted by the parametric models and that of the corresponding SPH
galaxy at $z=1$ (left) and $z=0.5$ (right). 
Each curve stands for a different parametric
model, and the
curves are normalised to unit integral.
}
\label{fig:sub3.11}
\end{figure}

Figure \ref{fig:sub3.11} compares the $r$-band mass-to-light ratio, $Y_r
\equiv M_*/L_r$ of SPH galaxies to that obtained from various parametric fits
to the SFH. The parametric fits are constrained to reproduce the $M_*$ of the
SPH galaxy. At both $z=0.5$ and $z=1$, the lin-exp model overestimates $Y_r$
for blue galaxies that have an increasing SFR at late times. Additionally,
for red galaxies with a truncated SFH, the lin-exp model underestimates $Y_r$
by $\sim$ 0.15 dex. The 2,3, and 4-parameter models can match both truncated
galaxy SFHs and the SFHs of galaxies with a rising SFR at late times,
yielding $Y_r$ values sharply peaked around the true $Y_r$ value. They fit 68\%
of galaxies to within 8\% at $z=1$, and within 13\% at $z=0.5$.


\section{Fitting Parametric Models to Galaxy Colours}

In dealing with observed galaxies, of course, we do not have a priori
knowledge of the SFH. Instead, the SFH and other physical
parameters of interest must be inferred from observables like the colours and
luminosity. In this Section, we test the efficacy of parametric SFH
models as practical tools by fitting them to the colours of galaxies
in our simulation and comparing these fits to the true SFHs.
We ignore the effects of dust extinction while computing the colours of both
SPH galaxies and parametric SFH models. While uncertainties in redshift,
dust extinction, and
metallicity can introduce additional errors and potential biases,
we focus in this work on the effect of the assumed
parametrisation of the SFH.

We use the FSPS stellar population synthesis code to compute the
luminosity of our simulated galaxies in five SDSS optical bands
($u$,$g$,$r$,$i$ and $z$), which gives us four colours. We fit our
previously described SFH models to these colours assuming a Gaussian
error on each colour of 0.02 magnitudes, typical of errors for
galaxies in the SDSS spectroscopic sample. While computing the colours,
we ignore
dust extinction and use the same SSPs to compute colours from the SFHs for
the SPH galaxy and the model fits, and therefore ignore the possibility of
template mis-match. Our fitting procedure is
based on $\chi^2$ minimization. We have implemented our 2, 3, and
4-parameter models as SFH options to FSPS and computed colours on a
grid of parameter values. We interpolate within this pre-computed grid
in our $\chi^2$ minimization procedure.

\begin{figure}
\centerline{
\epsfxsize=5.5truein
\epsfbox{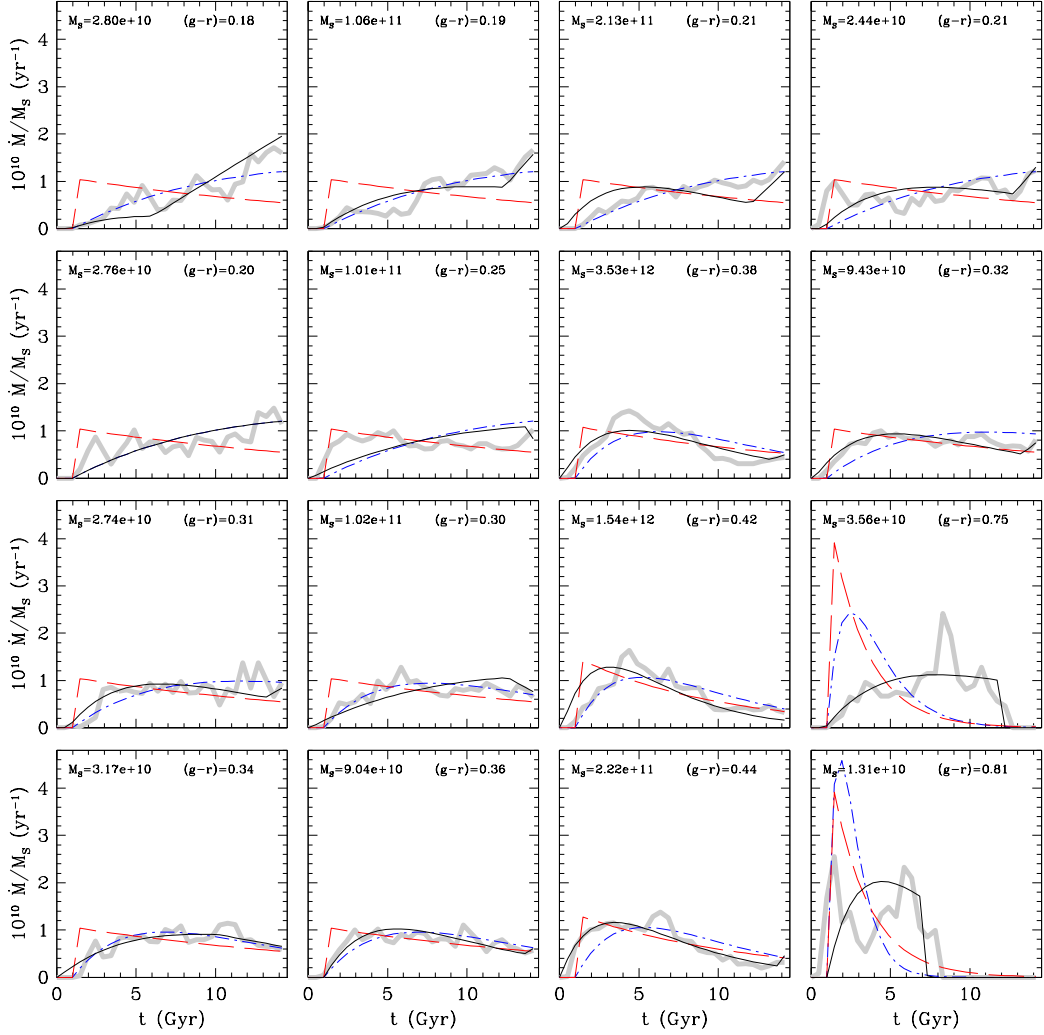} 
}
\caption{
Galaxy SFR versus time in the winds simulation. Each panel shows an
individual galaxy. We use the same set of galaxies as
Figure \ref{fig:sub3.1} but with models fit to the $z=0$ $ugriz$ colours
rather than
SFH. The thick gray curve shows the SFR in the simulation, the black
solid curve shows the best-fit 4-parameter model (see text), the blue
dot-dashed curve shows the best-fit lin-exp model, and the red dashed
curve shows the $\tau$-model.  }
\label{fig:sub4.1}
\end{figure}

Figure \ref{fig:sub4.1} shows the SFH of the same representative
sample of SPH galaxies illustrated in Figure~\ref{fig:sub3.1},
and gray bands are
repeated from that figure. Now, however, the model curves are found not
by minimizing the quantity in equation (\ref{eqm}) but by fitting the
four colours. As in Figure \ref{fig:sub3.1}, the solid black curve
shows the best-fit 4-parameter model, the dashed red curve shows the
best-fit $\tau$-model, and the blue dot-dashed curve shows the best-fit
lin-exp model. The top row shows blue galaxies, and each successive
lower row shows a redder colour. The first three columns from the left
show central galaxies in three mass bins, with mass increasing from
left to right, and the extreme right column shows satellite galaxies.
For most of our galaxies, the models are formally good fits to the
colours given our adopted 0.02 mag errors, though the $\tau$-model
sometimes fails to give a statistically acceptable fit for the
bluest galaxies.

In nearly all cases, the model fits reproduce the late-time SFH better
than the early SFH.  This is unsurprising, as the galaxy colour is
sensitive to late-time star formation but is minimally affected by
moderate shifts of age in the oldest stellar populations.
The best-fit $\tau$-model is never a good description of the true galaxy
SFH, showing the same generic discrepancies seen in
Figures~\ref{fig:sub3.1}-\ref{fig:sub3.3}.
The failures are worst for the bluest galaxies, where
the $\tau$-model cannot match the rising SFR at late times,
and for the reddest galaxies, where the $\tau$-model can produce
a truncated SFR at late times only with a strong burst of
very early star formation.
The lin-exp model provides a good description of
the SFH in many cases, for a variety of colours and SFH shapes.
The two examples where it fares badly are the satellites with
truncated SFHs (right column, two lower panels).  Like the
$\tau$-model, the single-parameter lin-exp model can only produce
very red colours by forcing rapid early star formation, while the
actual SFH of these galaxies is more extended before shutting
down at late times.  The 4-parameter model performs much better
than lin-exp for these galaxies, and it reproduces the rising
late-time SFR of the bluest galaxies.

For the Winds+Q population we see broadly similar trends, but
now the massive central galaxies also have a truncated SFH like
the two red satellites in Figure~\ref{fig:sub4.1}.
The fraction of galaxies for which the 4-parameter model
outperforms lin-exp is, therefore, larger.

\begin{figure}
\centerline{
\epsfxsize=5.0truein
\epsfbox{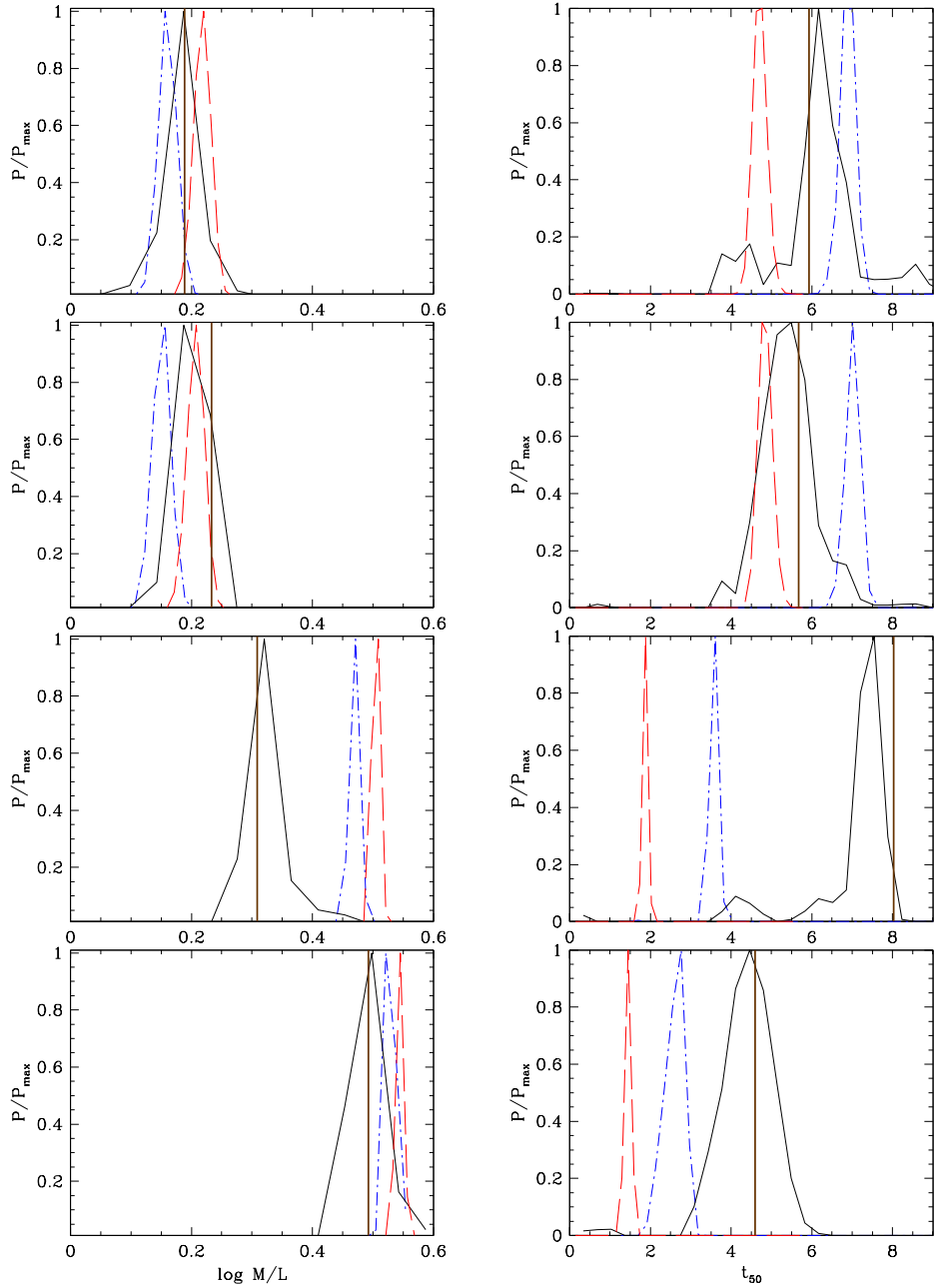} 
}
\caption{
Posterior probability distribution of the $r$-band mass-to-light ratio (left)
and t$_{50}$, the time at which 50\% of the stars are formed (right),
obtained by fitting parametric models to the $z=0$ galaxy $ugriz$
colours. Each row
stands for a galaxy. Each curve stands for a particular parametric
model - the 4-parameter model (solid black), the lin-exp model (blue
dot-dashed), and the $\tau$-model (red dashed). The grey solid line in
each panel shows the ``true" value of the mass-to-light ratio (left)
and t$_{50}$ (right) of the galaxy in the SPH simulation
}
\label{fig:subpdf}
\end{figure}

In practical applications, fits to observed SEDs are frequently
used to infer not the full SFH but high level physical parameters
such as stellar mass-to-light ratios (and corresponding stellar masses),
population ages, and current star formation rates.  Uncertainties
in these physical parameters (and, if desired, the covariance of
their errors) can be derived by marginalising over the parameters
of the fitted model.  Figure~\ref{fig:subpdf} illustrates this
approach for four of the simulated galaxies from
Figure~\ref{fig:sub4.1}, showing posterior probability distribution
functions (pdfs) of the mass-to-light ratio (left) and median population
age (right).
For the $\tau$-model and lin-exp model, we adopt a flat prior on $\tau$
over the range $0-20\Gyr$.
For the 4-parameter model we adopt the same prior on $\tau$,
a flat prior on $t_i$ over the range $0-1\Gyr$, a flat prior
on t$_{\rm trans}$ over the range $6-14.16\Gyr$
($t_0$ in our simulation), and a flat prior on $\theta = \tan^{-1}\Gamma$
such that the angle of the linear ramp can range uniformly from
$\theta = -\pi/2$ (instantaneous truncation) to $\theta = \pi/3$ 
(steeply rising).

The top two rows show the mass-to-light ratio and t$_{50}$ of
galaxies shown in rows 3 and 4, respectively, of column 3 of
Figure \ref{fig:sub4.1}. For both these galaxies, all three models fit
to the galaxy colours reproduce the SFH reasonably well.
The most probable $Y$ and $t_{50}$ for
the $\tau$-model and the lin-exp model
are reasonably close to the true SPH values, although the true
value is sometimes outside the formal 95\% confidence interval.
Because of its greater flexibility, the 4-parameter model allows
a larger range of $Y$ and $t_{50}$, but
the peak of the posterior pdfs are very close to the true values,
and the true values are always within the 68\% confidence
interval. The bottom two panels show galaxies whose SFH is truncated
(rows 3 and 4 of column 4 in Figure \ref{fig:sub4.1}). As discussed
earlier, both the $\tau$-model and the lin-exp model fail to match the
truncated SFH, instead putting too much star formation at early times to
match the SPH colours. Because they overpredict the age of
the stellar population, they overpredict the mass-to-light ratio. In
contrast, the 4-parameter model, which matches the truncation in the
SFH, predicts a posterior probability distribution for the
mass-to-light ratio whose peak is remarkably close to the true value
for both galaxies. The true value of t$_{50}$ is within the
4-parameter model predicted 95\% confidence interval in one case
(third row), and is very close to the peak of the posterior
probability distribution in the other (bottom row).

\begin{figure}
\centerline{
\epsfxsize=5.0truein
\epsfbox{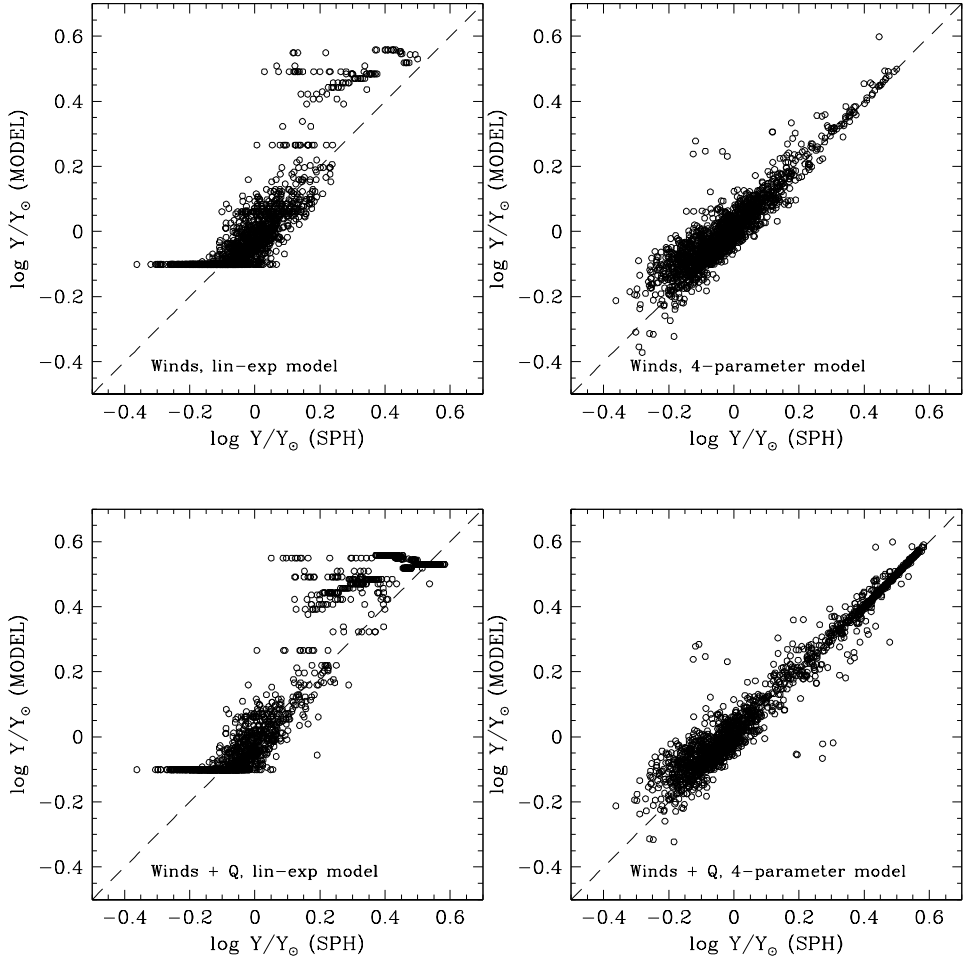} 
}
\caption{
Mode of the posterior probability distribution of the mass-to-light
ratio from parametric model SFH fits to $z=0$ $ugriz$ colours versus
mass-to-light
ratio of SPH galaxies. Each point is an individual galaxy. There are
1,828 galaxies plotted in the upper panels and 1,723 in the lower panels. The
left panels show
the lin-exp model and the right panels the 4-parameter model. The upper
panels are fits to galaxies
in the Winds population and the bottom panels to those in the Winds+Q
population.}
\label{fig:sub4.2}
\end{figure}

When investigating statistics for a large population of galaxies
(e.g., the galaxy stellar mass function), it is common practice
to take the best-fit model parameters for each individual galaxy,
though a more sophisticated analysis could consider the full
posterior pdf on a galaxy-by-galaxy basis.  In what follows we
will take the ``best-fit'' value of a parameter to mean the
mode of that parameter's posterior PDF.
Each panel of Figure~\ref{fig:sub4.2} plots the best-fit
stellar mass-to-light ratio from a lin-exp (left) or 4-parameter (right)
model fit to the colours of individual SPH galaxies against the galaxies'
true mass-to-light ratios.  The mass-to-light ratios in the SPH
Winds population (top row) range from $\sim 0.4 Y_\odot$ to
$\sim 3Y_\odot$, where $Y_\odot = 1 M_\odot/L_\odot$.
For the bluer galaxies, with $Y < 2Y_\odot$, the lin-exp model
predicts $Y$ quite accurately, though SPH galaxies with steeply
rising late-time SFR have mass-to-light ratios lower than the
minimum value $Y \approx 0.8 Y_\odot$ that the lin-exp model
can produce (with $\tau$ forced to its limiting value of 20 Gyr).
The behaviour for the red galaxies with a truncated SFR is more problematic.
Because lin-exp does not allow sharp truncation, it attempts
to produce red colours by forcing star formation very early
(see the lower right panels of Fig.~\ref{fig:sub4.1}).  As a result,
the lin-exp fits overpredict the true $Y$ for these galaxies.
The 4-parameter model yields a good correlation between the
best-fit $Y$ and the true value across the full range, with only
a small number of outliers.  The performance for the Winds+Q
population (lower panels) is similar in both cases, but now the
fraction of ``red and dead'' galaxies at large $Y$ is higher
because it includes massive centrals.

\begin{figure}
\centerline{
\epsfxsize=4.0truein
\epsfbox{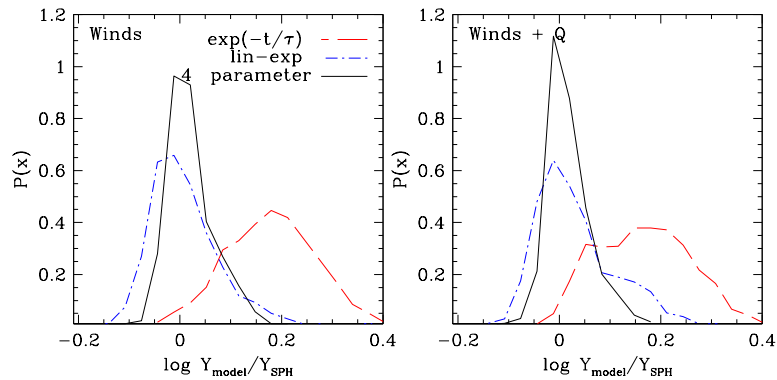} 
}
\caption{
Distribution of differences between the $r$-band mass-to-light ratio
predicted by the parametric model SFH fits to $z=0$ $ugriz$ colours and that
of the corresponding SPH
galaxy. Each curve stands for a different parametric model, and the
curves are normalised to unit integral.
}
\label{fig:sub4.4}
\end{figure}

Figure \ref{fig:sub4.4} shows the distribution of the differences
between the mass-to-light ratios obtained from fitting different
parametric models to the optical colours of galaxies and their
mass-to-light ratios in the simulation. The $\tau$-model typically
overestimates the mass-to-light ratio by a factor of $\sim$1.5, but
the error can be as high as a factor of $\sim$ 3. The lin-exp model
generally does better, but it makes significant errors in either
direction.  Most notably, as already seen in Figure~\ref{fig:sub4.2},
it overpredicts $Y$ for the reddest galaxies, producing the tail at
high $Y_{\rm model}/Y_{\rm SPH}$ in Figure~\ref{fig:sub4.4}.
The 4-parameter model estimate for the mass-to-light ratio
is within 10\% of the true value for 68\% of galaxies in the Winds and
Winds+Q populations, though it shows a weaker version of the same
asymmetry seen for lin-exp.

\begin{figure}
\centerline{
\epsfxsize=5.0truein
\epsfbox{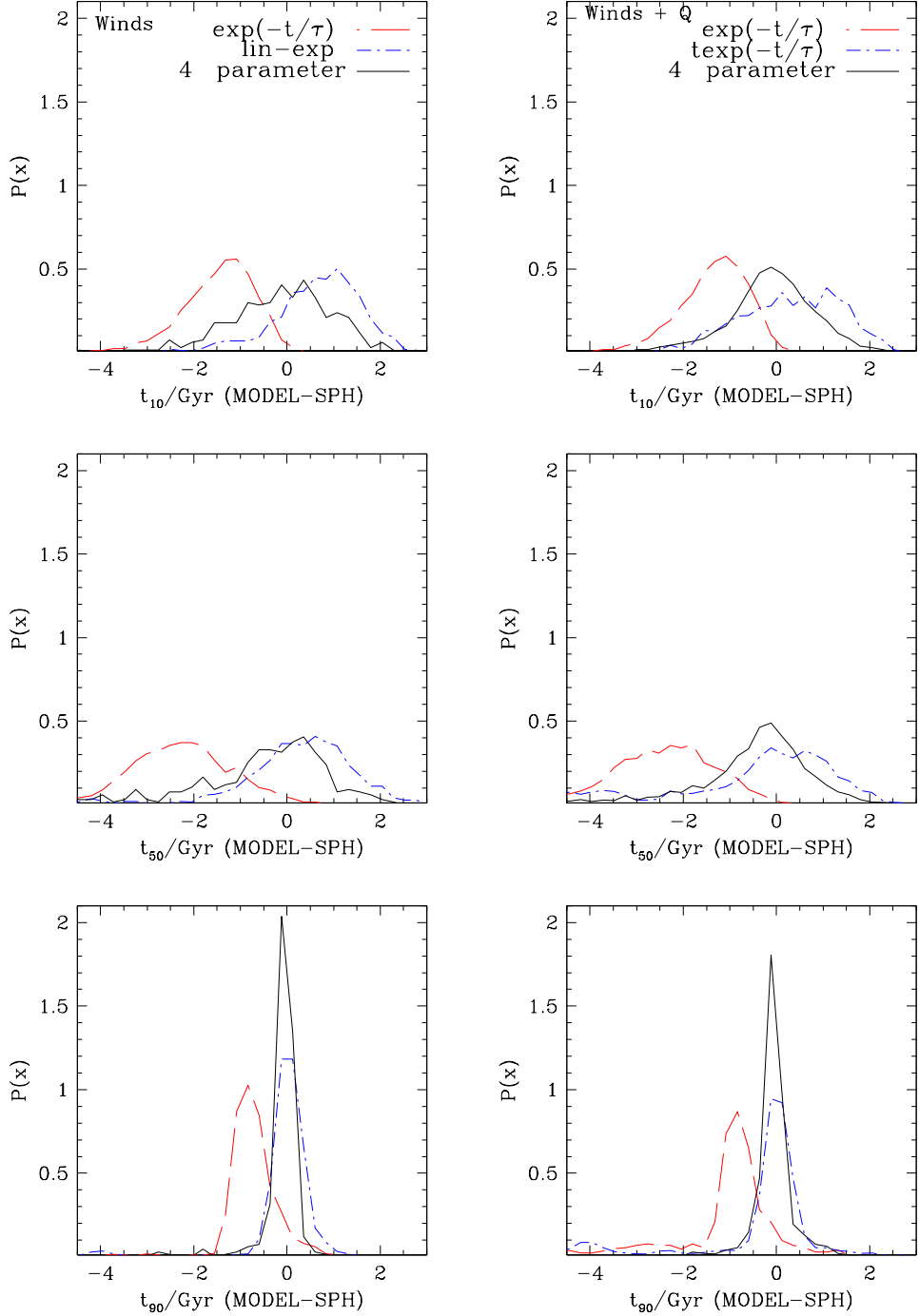} 
}
\caption{
Distribution of differences between age of the galaxy stellar
population predicted by the parametric model SFH fits to $z=0$ $ugriz$
colours and their age in the
winds population (left) and the Winds + Q population (right). Each curve
stands for a different parametric model and the curves are normalised
so that the area under each curve integrates to unity. $t_{10}$, $t_{50}$ and
$t_{90}$ stand for the time at which 10\%, 50\% and 90\% of the
stars were formed respectively.
}
\label{fig:sub4.5}
\end{figure}

Figure~\ref{fig:sub4.5} presents a similar analysis for stellar population
ages, showing the distribution of differences between model fit
values and SPH galaxy values for the times when 10\% ($t_{10}$, top row),
50\% ($t_{50}$, middle row), and 90\% ($t_{90}$, bottom row) of the
stars have formed.  The 4-parameter model correctly predicts
$t_{10}$ and $t_{50}$ to within 1 Gyr and $t_{90}$ to within
0.3 Gyr for 68\% of galaxies in both the Winds and Winds+Q populations,
and it shows no significant bias, though the tails of the difference
distribution are slightly asymmetric. The distribution for lin-exp is
qualitatively similar, but it is biased
towards high $t_{10}$ and $t_{50}$ by $0.5-1\Gyr$, and the distribution
for $t_{90}$ is less sharply peaked.  The $\tau$-model fits are systematically
biased towards older population ages (smaller $t_{10}$, $t_{50}$, and
$t_{90}$), by $1-2\Gyr$ for $t_{10}$ and $t_{50}$ and by $1\Gyr$
for $t_{90}$.

\begin{figure}
\centerline{
\epsfxsize=5.0truein
\epsfbox{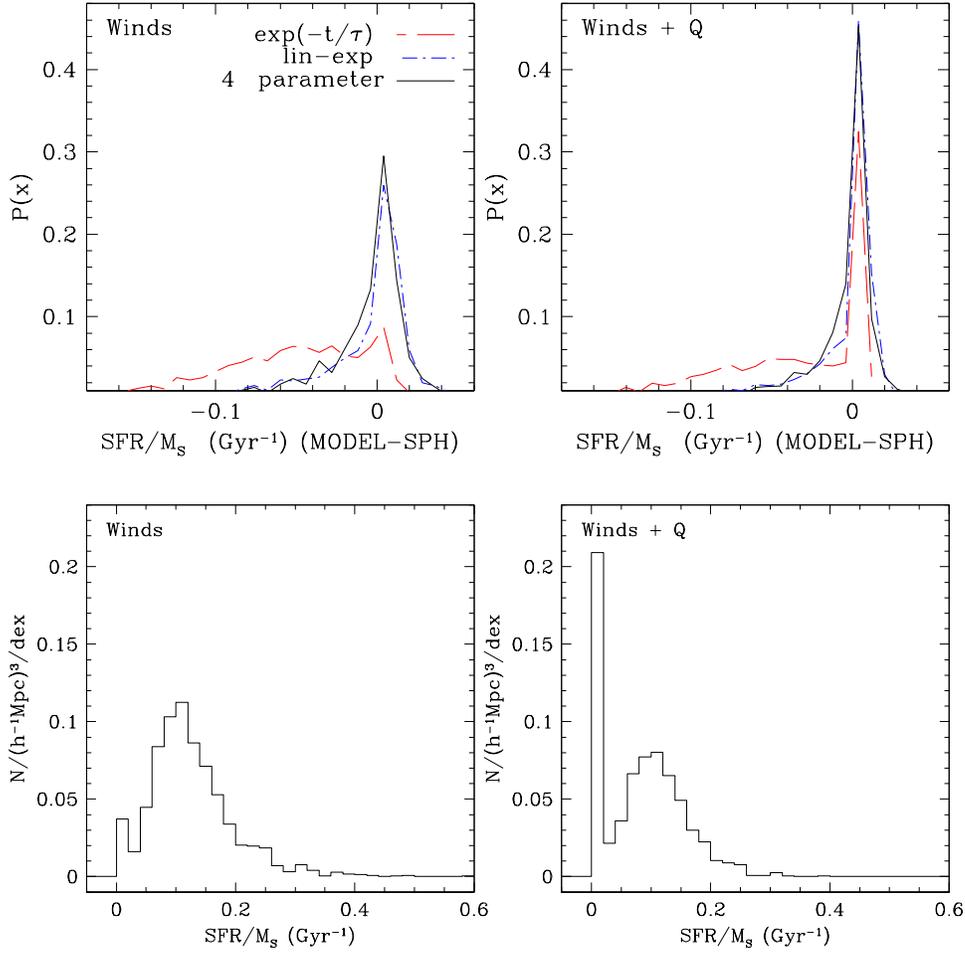} 
}
\caption{
Distribution of differences between the specific star formation rate
(sSFR) predicted by the parametric model SFH fits to $z=0$ $ugriz$ colours
and the Winds
population (top left) and the Winds + Q population (top right). Each curve stands
for a different parametric model, and the curves are normalised to
unit integral.  Histograms in the two bottom panels show the distribution of sSFR in the
two galaxy populations. 
}
\label{fig:sub4.6}
\end{figure}

Figure \ref{fig:sub4.6} shows similar results for specific star
formation rates (sSFR$\,\equiv \dot{M}_*/M_*$) at $z=0$.
For context, inset panels show the histograms of sSFR in the
two galaxy populations.
Since sSFR is strongly correlated with colour at $z=0$, models fit to the
colours generally reproduce the sSFR quite accurately.
However, the $\tau$-model is unable to match the colours of
galaxies with rising late-time SFR, and it consequently underestimates
their sSFR.  All models successfully reproduce low sSFRs for the
reddest galaxies, so the peak at near-perfect agreement is higher
in the Winds+Q population, where the proportion of such galaxies
is larger.  {\it Fractional} errors in the sSFR can be large when
the value is extremely small, but for most purposes it is the
absolute error that is more relevant.
We have carried out the same analyses shown in 
Figures~\ref{fig:sub4.4}-\ref{fig:sub4.6} at $z=0.5$ and $z=1$,
for the same rest-frame colours.
While we do not show the plots here, the trends are similar,
with the 4-parameter model producing moderate improvements
over lin-exp and substantial improvements over the $\tau$-model
in recovering stellar mass-to-light ratios, population ages,
and sSFRs.
We have also checked that using the 2-parameter model 
with our recommended choices of $t_i$ and $\ttr(z)$ yields
similar results to those of the 4-parameter model.

The $\tau$-model, which enforces declining SFHs, fails to match the
mass-to-light ratios and stellar population ages of blue galaxies with
ongoing star formation, often over predicting the ratios by a factor
of $\sim$ 2. The lin-exp model performs better, providing more precise
estimates, but it is often biased. In contrast, the 4-parameter model
matches the physical parameters of SPH galaxies quite well. In the
4-parameter model, the individual parameters have partially degenerate
effects on the SFH, but this degeneracy does not degrade the
determinations of these physical quantities, since a similar SFH
implies similar quantities and a similar fit to the data regardless of
what parameter combinations produce it.  With limited data (e.g., two
or three colours, or large colour errors), individual model parameters
may be poorly determined, but physical quantities may still be well
constrained after marginalization.

\begin{figure}
\centerline{
\epsfxsize=4.0truein
\epsfbox{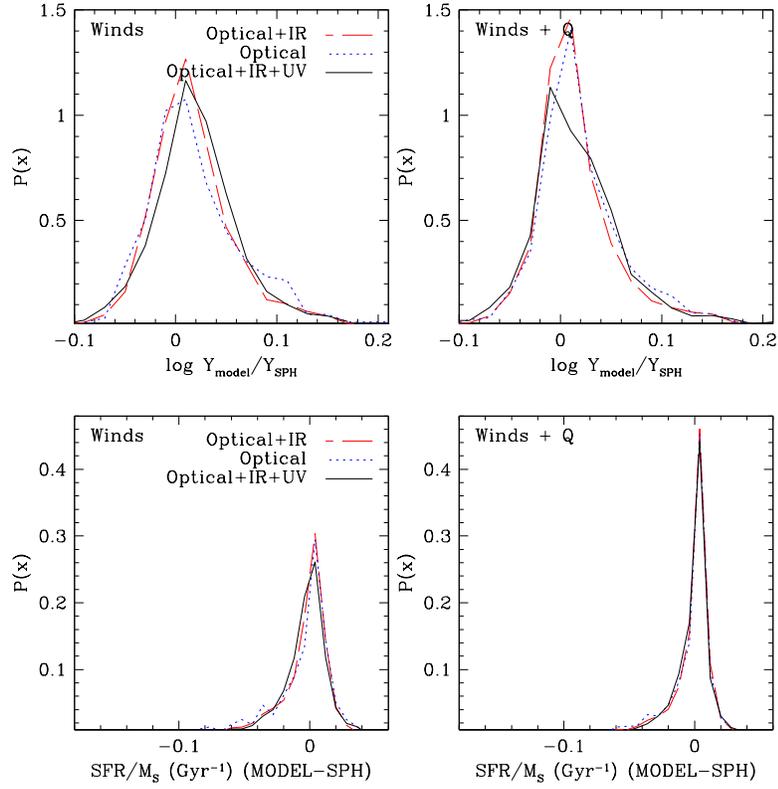} 
}
\caption{
Impact of adding IR or IR+UV colours to optical colours when
inferring the $r$-band mass-to-light ratio (top) or sSFR (bottom).
Curves show the distribution of errors from fits of the
4-parameter model using optical colours only (dotted),
optical+IR (dashed), or optical+IR+UV (solid).
}
\label{fig:sub5.1}
\end{figure}

So far, we have only considered optical colours. We have also examined the
effects of adding IR and UV colours to the optical data. Specifically,
we compute SPH galaxy fluxes in the 2MASS J, H, and K bands and the
GALEX NUV and FUV bands, and fit models to the combined data sets
again assuming an error of 0.02 magnitudes on each colour.
Figure \ref{fig:sub5.1} shows the distribution of errors
in $Y$ and sSFR from fits of the 4-parameter model to optical colours
alone, optical+IR colours, and optical+IR+UV colours.
Somewhat surprisingly, adding IR and UV colours does not noticeably
reduce the scatter in recovering $Y$ or sSFR; at least with regard
to our 4-parameter model, the optical colours already contain all
of the relevant information.  We find similar results for population ages.
Note, however, that we have not included dust extinction or metallicity
in our models, and we have assumed that galaxy redshifts are known
so that the rest-frame colours are available.  Since optical colours already
suffice to constrain the SFH in our framework,
the additional information in IR and UV data can be
applied to constrain extinction, metallicity, and (if necessary) redshift.
Broad wavelength coverage is especially important in photometric
redshift studies, as UV and IR data help to unambiguously identify
breaks in the SED.

\section{Discussion}

The star formation histories (SFHs) of galaxies in our SPH simulations
are generally smooth, governed by the interplay between cosmological
accretion and star-formation driven outflow.  As a result, they can be
well described by models with a small number of free parameters, and
this remains true after we implement a post-processing quenching
scheme designed to reproduce the observed red colours and stellar mass
distributions of the central galaxies in massive halos.  While the
simulations are far from perfect, they include many realistic aspects
of cosmological growth, gas dynamics and cooling physics, and
feedback.  They can, therefore, provide guidance to the classes of
models that are most useful for fitting observed galaxy populations.

One of the models most commonly used for this purpose, the
exponentially decaying ``$\tau$-model'', gives a quite poor
representation of our simulated galaxies because of its implicit
assumption that star formation is most rapid at the earliest epochs
and declines thereafter.  Adding an initial burst to a $\tau$-model
would only exacerbate this problem, and allowing a start at $t_i > 0$
helps but only moderately.  Fitting the colours of our simulated
galaxies with a $\tau$-model (with start time $t_i = 1\,$Gyr) leads to
inferred stellar mass-to-light ratios that are systematically too high,
by a typical factor $\sim 1.5$, and to inferred stellar population ages
that are too large, typically by $1-2\,$Gyr.  Inferred specific star
formation rates (sSFRs) can be either too high or too low, with
substantial scatter about the true values.

The \linexp\ model, with $\dot{M}_* \propto
(t-t_i)\exp[-(t-t_i)/\tau]$, gives a much better description of the
time profiles of star formation in our simulation.  We find little
star formation in our simulations before $t=1\,$Gyr, reflecting the
time required to build up massive systems that can support vigorous
star formation, so the model is improved by setting $t_i=1\,$Gyr
instead of $t_i=0$.  Fitting the \linexp\ model to the optical colors
of SPH galaxies largely removes the biases in $M/L$ ratios and
population ages that arise with $\tau$-model fitting, and it reduces
the scatter between the inferred and true values for these quantities and
for sSFRs.  If one is going to fit a galaxy SFH with a one-parameter
model, the \linexp\ model with $t_i=1\,$Gyr is the one to choose.

The shortcoming of the \linexp\ model is that it ties late-time star
formation to early star-formation: a rapid early build-up (short
$\tau$) necessarily implies a low sSFR at low redshift.  Early and
late star formation are correlated in SPH galaxies, but they are not
so perfectly correlated that galaxies lie on a 1-parameter family of
SFH.  Our 2-parameter model avoids this problem by changing
from \linexp\ to a linear ramp after $\ttr = 10.7\gyr$, decoupling
early and late evolution.  For many galaxies, this additional freedom
makes little difference, but the two-parameter model offers a
significantly better description of the bluest galaxies, which have
rising star formation rates at late times, and of the reddest
galaxies, which have truncated star formation.  Fitting galaxy colors
with the 2-parameter model removes the small systematic biases in
$M/L$, population ages, and sSFRs that remain with \linexp\ fitting,
and it reduces the scatter between the true and fitted values.

Our 3-parameter model turns $\ttr$ into a fit parameter instead of
fixing it at 10.7$\gyr$ (or more generally 75\% of the current cosmic time), and our 4-parameter model additionally turns
the start time $t_i$ into a fit parameter.  This additional freedom
only marginally improves the description of SPH galaxy SFHs or the
accuracy of inferred parameter values.  However, it avoids hard-wiring
these ages into the model, and it provides some safeguard against the
possibility that they are too strongly tied to the specifics of our simulation.
For example, the preference for $t_i \approx 1\gyr$ could be affected
by our mass resolution. In spot checks on a simulation with the same volume
but $8\times$
higher mass resolution, which became available after we had completed
most of our analysis, we find that all of our results for galaxy
SFHs continue to hold, but the best fit value for $t_i$ shifts
slightly, from 1 Gyr to 0.83 Gyr.

Our recommended strategy, therefore, is to adopt the 4-parameter model
for fitting galaxy colours or SEDs and marginalize over model
parameters when computing physical quantities of interest such as
$M/L$ ratios, population ages, and specific star-formation rates.
To enable this approach, we have added the 4-parameter model as an
option to the FSPS population synthesis code
\citep{conroy09}.\footnote{Publicly available at
{\tt http://code.google.com/p/fsps/}.}
In addition to colours,
the FSPS code can compute full galaxy spectra
for fitting to spectroscopic data.
In place of marginalization, a less laborious but less robust strategy is to
estimate physical quantities from the best-fit 2-parameter model,
with $t_i$ and $\ttr$ fixed to our recommended fiducial values.
For our SPH galaxies, this
procedure actually yields a {\it better} match between inferred and
true quantities because the adopted priors on $t_i$ and $\ttr$ are a
good match to the simulations.  However, these priors may be overly
strong for fitting real galaxies, and the marginalization approach
with the 4-parameter model is more conservative.

When fitting the $ugriz$ colours of our SPH galaxies at $z=0$,
assuming 0.02 mag colour errors, we are able to determine $r$-band
mass-to-light ratios with typical errors of $\pm 13\%$ (the range
encompassing 68\% of simulated galaxies).  The corresponding error for
median population age is 0.9 Gyr, and $t_{90}$, the time by which 90\%
of stars form, has a smaller error of 0.3 Gyr.
Adding near-UV or near-IR colors produces little further improvement
because these quantities are already well determined given the assumed
$ugriz$ errors, and additional wavelengths do little to break the
degeneracies in the 4-parameter model.  When fitting real galaxies,
one would also need to include dust extinction as an additional
parameter with an assumed extinction law (or marginalizing over a
range of extinction laws).  Including dust extinction will only
moderately increase $M/L$ uncertainties because reddening induced by
dust or by increased stellar population age has a similar impact on
$M/L$ \citep{bellanddejong}.  Conversely, dust extinction increases
sSFR uncertainty because increasing dust or increasing late-time star
formation have opposite effects on galaxy colours.  In the absence of
spectroscopic redshifts, one also needs to fit for galaxy photo-$z$
along with the stellar population quantities.  In this situation, UV or
near-IR data may play a more critical role, breaking degeneracies
among SFH, extinction, and redshift. However, we caution that \cite{taylor11}
find that SPS models do not provide good fits to the full optical-to-NIR SEDs
of the galaxies they observe, possibly indicating inconsistencies between the 
SED shapes of real galaxies and those of the models.

We have extended our analysis to higher redshifts,
finding that at $z=0.5$ and $z=1$ the \linexp\ model suffers
from similar shortcomings as at $z=0$, which are overcome by models that
decouple the early and late SFRs. Our 2-parameter model, with $t_i$ = 1 Gyr
and $\ttr$ scaled by the age of Universe, provides a significantly better
description of the SFH of SPH galaxies.  As at $z=0$, 
allowing $t_i$ and $\ttr$ to be free parameters only marginally improves
the description of SPH galaxy SFHs.
At all redshifts, one should bear in mind that our simulation models 
galaxies with $M_* > 10^{10} M_\odot$, and that the SFHs that 
characterize much lower mass galaxies could be different both 
in overall form and in the level of stochasticity.

Sources of uncertainty in the SED fitting technique at even higher redshifts
have been investigated by other authors. \cite{lee10} apply standard SED
fitting techniques to infer the physical parameters of Lyman Break Galaxies at
$z \sim 3.4 - 5$ in a mock catalogue constructed from semi-analytic models
of galaxy formation,  finding that SFRs are systematically underestimated
and mean stellar population ages overestimated because of differences
between the galaxy SFHs predicted by their semi-analytic models and the
$\tau$-model SFH assumed in their SED fitting technique. Because of the mass
resolution of our simulation, the SFHs of $z \ge 2$ galaxies are too noisy
to allow us to carry out a direct comparison, but our results at $z \le 1$
are qualitatively similar to their high $z$ results, highlighting similar
discrepancies in the commonly used $\tau$-model SFH.

Fitting the low-order parametrised models presented here should be
more precise than fitting general stepwise SFHs.  In essence, one is
imposing a prior of approximate continuity to extract more from the
data and reject pathological fits.  This approach may also be more
robust to uncertainties in the population synthesis models, since the
strong spectral features that appear in stellar populations at
specific ages may lead to artificial features in stepwise SFH fits.
However, it is possible that the SFHs of real galaxies are more
complex than those of our simulated galaxies, with bursts playing a
more important role or truncation followed by rejuvenation.  It will
be interesting to search for evidence of such deviations to better
constrain the potential contribution of ``punctuated'' star formation
in galaxies of different stellar mass or morphology, or even in
individual components of galaxies.  These searches can be best carried
out with full spectroscopic data rather than with colors alone, or
better yet, with resolved stellar populations in nearby galaxies.  

\section * {ACKNOWLEDGEMENTS}
We acknowledge support from NASA ATP grant NNX10AJ95G.

\section * {APPENDIX: COMPARISON WITH BEHROOZI SFH PARAMETRISATION}

\cite{behroozi13} advocate a parametrisation of the SFH based on reconstruction
of average SFHs using observed galaxy stellar mass functions, specific star
formation rates and cosmic star formation rates. The functional form they
advocate is given by:
\begin{equation}
\rm{SFR}(t) = A~ [~ (t/{\tau})^{B} + ~(t/{\tau})^{-C}]^{-1}.
\label{eq:behroozi}
\end{equation}

This model contains three free parameters, $\tau$, $B$ and $C$, in addition to
the overall normalization $A$. We fit this model to the SFHs of SPH galaxies
allowing $B$ and $C$ to vary between 0 and 25.

\begin{figure}
\centerline{
\epsfxsize=5.0truein
\epsfbox{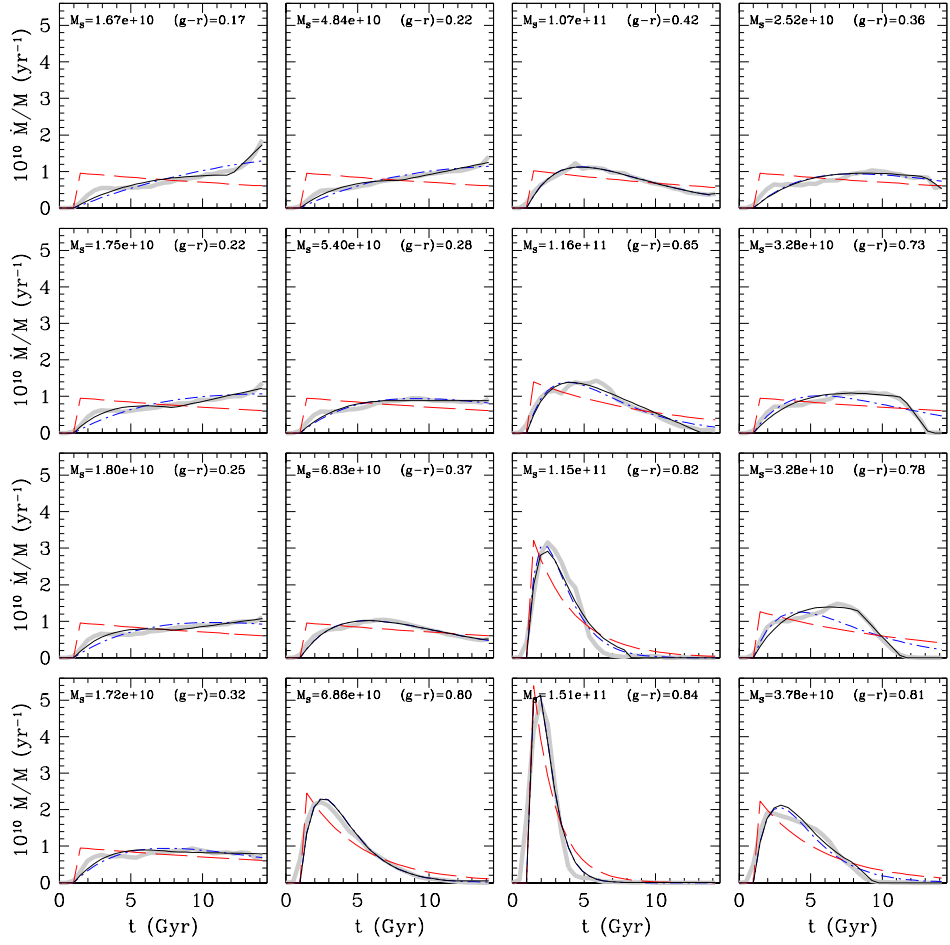} 
}
\caption{
Same as Figure 4, but for the Winds + Q population.  }
\label{fig:sub3.3}
\end{figure}

\begin{figure}
\centerline{
\epsfxsize=5.0truein
\epsfbox{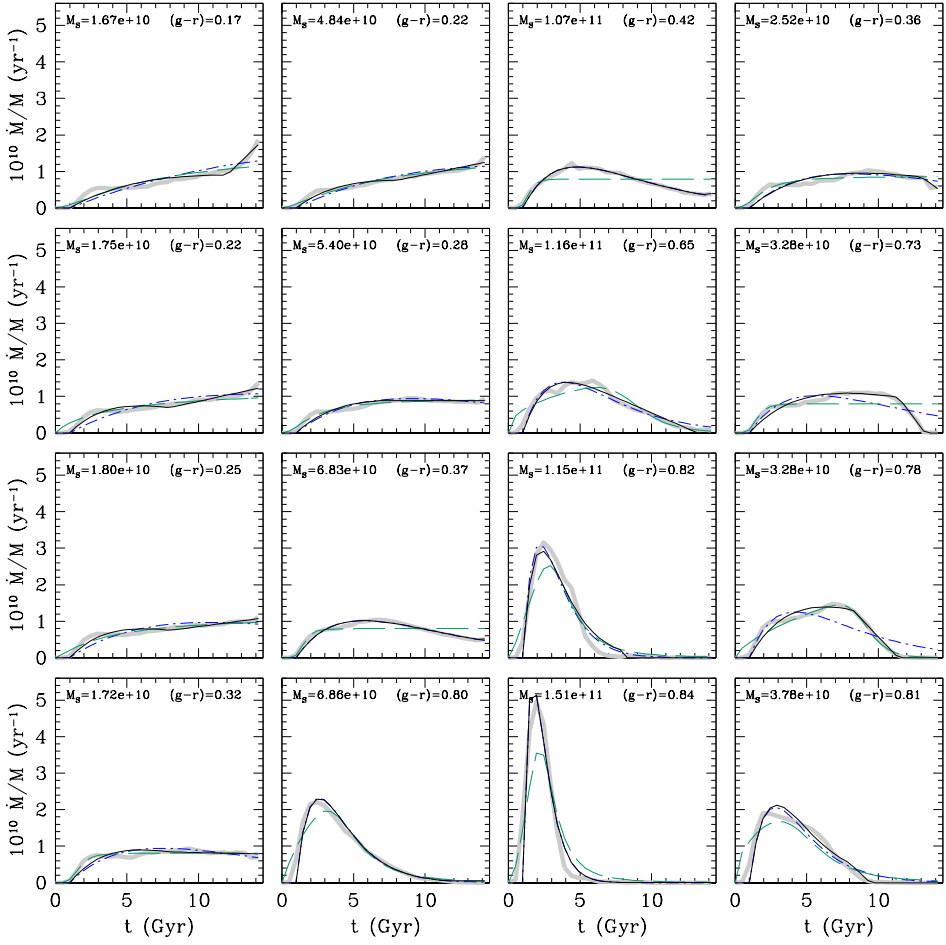}
}
\caption{
Same as Figure \ref{fig:sub3.3}, but showing the \protect\cite{behroozi13} SFH
parametrisation (long dashed green) fit to the average SFH of galaxies in bins of
mass and colour in our Winds + Q population. For comparison, we also show
the lin-exp model (blue dot dashed) and 4-parameter model (black solid) fits. 
}
\label{fig:sub16}
\end{figure}

Figure \ref{fig:sub16} shows the result of fitting this model to the average
SFH of galaxies in the Winds + Q model in bins of mass and colour. We show
the same set of SFHs as Figure \ref{fig:sub3.3}. For most galaxies, this 
model provides a good description of the early SFH. However, because it
ties the late time SFR to the early SFR, it does not adequately match the
SFH of galaxies with a rising SFR at late times such as those in the top
two rows of the left most column. SFHs that are flat or gradually declining
at late times are generally well described by the B-model, although for a
significant fraction of galaxies, the B-model fails to match the late time SFH.
But unlike the \linexp\ model, the \cite{behroozi13} parametrisation can
match truncated SFHs well by employing a large value of $B$, and a value
of $C$ close to 0. The B-model generally provides a substantially better
description of the SFH of SPH galaxies than the $\tau$-model or the \linexp\ 
model, but it does not perform as well as the 4-parameter model.

In practice, our 2-parameter model fits the SFH of SPH galaxies nearly
as well as our 4-parameter model, and significantly better than equation
\ref{eq:behroozi} in situations where they disagree.  Thus, the better
performance of our model owes to its functional form that decouples early
and late-time star formation, and not to the number of parameters.

\clearpage
\onecolumn


\clearpage

\clearpage
\bibliographystyle{mn2e}



\end{document}